\documentclass[conference]{IEEEtran}
\IEEEoverridecommandlockouts
\usepackage{cite}
\usepackage{amsmath,amssymb,amsfonts}
\usepackage{mathtools}
\mathtoolsset{showonlyrefs}
\usepackage{bm}
\usepackage{graphicx}
\usepackage{textcomp}
\usepackage{xcolor}
\usepackage{booktabs}
\usepackage{algpseudocode}
\usepackage{algorithmicx}
\usepackage[shortlabels]{enumitem}
\usepackage{multicol}
\usepackage{tablefootnote}
\usepackage{multirow}

\let\originalleft\left
\let\originalright\right
\renewcommand{\left}{\mathopen{}\mathclose\bgroup\originalleft}
\renewcommand{\right}{\aftergroup\egroup\originalright}

\def\BibTeX{{\rm B\kern-.05em{\sc i\kern-.025em b}\kern-.08em
    T\kern-.1667em\lower.7ex\hbox{E}\kern-.125emX}}
\begin{document}

\newif\ifanonymous
\anonymousfalse

\title{Double-Precision Matrix Multiplication Emulation via Ozaki-II Scheme with FP8 Quantization}

\ifanonymous
 \author{Anonymous Author(s)}
\else
\author{%
    \IEEEauthorblockN{Yuki Uchino}
    \IEEEauthorblockA{%
        \textit{Center for Computational Science}\\
        \textit{RIKEN}\\
        Hyogo, Japan\\
        yuki.uchino.fe@riken.jp%
    }
    \and
    \IEEEauthorblockN{Katsuhisa Ozaki}
    \IEEEauthorblockA{%
        \textit{Department of Mathematical Science} \\
        \textit{Shibaura Institute of Technology}\\
        Saitama, Japan\\
        ozaki@sic.shibaura-it.ac.jp%
    }
    \and
    \IEEEauthorblockN{Toshiyuki Imamura}
    \IEEEauthorblockA{%
        \textit{Center for Computational Science}\\
        \textit{RIKEN}\\
        Hyogo, Japan\\
        imamura.toshiyuki@riken.jp%
    }
}
\fi

\maketitle

\begin{abstract}
In this paper, we propose a method for emulating double-precision general matrix--matrix multiplication (DGEMM), a fundamental and performance-critical kernel in many high-performance computing applications.
Ozaki-I and Ozaki-II are established DGEMM emulation schemes via low-precision matrix multiply-accumulate (MMA) units.
For the Ozaki-I scheme, INT8-, FP8-, and FP16-based implementations have been proposed, all of which can be realized based on the same underlying algorithmic structure.
In contrast, although INT8-based implementations of the Ozaki-II scheme have been reported, the original algorithm cannot be directly adapted to exploit FP8 MMA units.
In several recent architectures, such as NVIDIA Blackwell Ultra and NVIDIA Rubin, INT8 performance has been reduced, making reliance on INT8 alone insufficient. 
Therefore, we introduce a novel technique to demonstrate DGEMM emulation based on the Ozaki-II scheme that operates on FP8 MMA units.
Compared to the FP8-based Ozaki-I scheme, our method significantly reduces the computational cost and enables efficient FP64 emulation.
\end{abstract}


\section{Introduction}
\label{sec:Introduction}

\subsection{Background}
In high-performance computing (HPC) applications, FP64 arithmetic remains a fundamental building block for ensuring sufficient numerical accuracy and stability.
However, recent computing architectures, driven by strong market demand for accelerating rapidly expanding AI workloads, have achieved exceptionally high performance and energy efficiency in low-precision arithmetic. 
In particular, low-precision matrix multiply-accumulate (MMA) units designed for matrix operations have improved rapidly across hardware generations. 
As a result, low-precision arithmetic formats such as FP4, FP6, FP8, INT8, FP16, and BF16 now provide computational throughput that far exceeds that for conventional FP32 and FP64 arithmetic, for which performance improvements have been limited. 
In recent architectures, FP32 and FP64 arithmetic units are either insufficiently provisioned or assigned significantly lower design priority. 
Under such hardware trends, achieving sustained performance improvements for HPC applications increasingly requires mixed-precision algorithms and emulation techniques that effectively leverage low-precision arithmetic to realize high-precision computations.

In recent architectures, such as NVIDIA Blackwell Ultra and NVIDIA Rubin, INT8 computational resources have been substantially reduced, with greater emphasis placed on low-precision floating-point arithmetic such as FP4 and FP8.
This architectural shift is clearly reflected in the peak dense throughput of recent NVIDIA GPUs (Table~\ref{tab:gpu_peak_performance}).
While the NVIDIA B200 Blackwell GPU provides comparable throughput for FP8 and INT8 operations, the B300 Blackwell Ultra GPU and the Rubin GPU exhibit a drastic reduction in INT8 performance.

\begin{table}[htbp]
\centering
\caption{Single-GPU specifications of recent NVIDIA GPUs~\cite{Rubin,Hopper,Blackwell}\label{tab:gpu_peak_performance}}
\begin{tabular}{@{ }c@{\quad }c@{\quad }c@{\quad }c@{\quad }c@{\quad }c@{ }}
\toprule
     & \multicolumn{2}{c@{\quad }}{Blackwell} & \multicolumn{2}{c@{\quad }}{Blackwell Ultra} & Rubin \\
     & B200 SXM & GB200    & B300 SXM & GB300         & \\
\midrule
FP4 (TFLOP/s)     & 9000  & 10000  & 14000  & 15000  & 35000  \\
FP6 (TFLOP/s)     & 4500  & 5000   & 4500   & 5000   & 17500  \\
FP8 (TFLOP/s)     & 4500  & 5000   & 4500   & 5000   & 17500  \\
INT8 (TOP/s)      & 4500  & 5000   & 150    & 166    & 250    \\
FP16 (TFLOP/s)    & 2250  & 2500   & 2250   & 2500   & 4000   \\
BF16 (TFLOP/s)    & 2250  & 2500   & 2250   & 2500   & 4000   \\
FP32 (TFLOP/s)    & 75    & 80     & 75     & 80     & 130    \\
FP64 (TFLOP/s)    & 37    & 40     & 1.2    & 1.4    & 33     \\
Bandwidth (TB/s)  & 7.7   & 8      & 7.7    & 8      & 22     \\
\bottomrule
\end{tabular}
\end{table}

This trend is not universal and INT8 capability remains important on many currently deployed accelerators. 
For example, NVIDIA B200 and earlier GPUs~\cite{A100, H100, Hopper, Blackwell}, as well as AMD Instinct MI355X and earlier hardware~\cite{MI300A, MI300X, MI325X, MI350X, MI355X}, provide comparable peak throughput for FP8 and INT8 matrix operations (based on vendor specifications). 
In addition, some accelerators continue to prioritize INT8 or do not publicly expose FP8 support. 
For instance, the Intel NPU documentation for Core Ultra 200H/200U~\cite{Intel200H200U} lists INT8 and FP16, but not FP8, as supported inference precisions for internal primitives.
Likewise, Google Cloud TPU v6e is documented to have 1836~TOP/s INT8 peak compute~\cite{TPUv6e}, while a Google Cloud TPU7x cross-generation comparison table lists 918~TFLOP/s FP8 peak compute for v6e~\cite{TPU7x}, indicating a nominal 2$\times$ INT8-to-FP8 ratio.
These examples suggest that the relative balance between INT8 and FP8 remains hardware-dependent.

\subsection{Problem Statement and Related Work}
In this study, we focus on double-precision (FP64) general matrix--matrix multiplication (DGEMM), which is an indispensable and performance-critical computational kernel in a wide range of HPC applications, and investigate its emulation using low-precision arithmetic.
Two representative approaches for DGEMM emulation are the Ozaki-I scheme~\cite{ozaki2012error,ozaki2013generalization} and the Ozaki-II scheme~\cite{ozaki-scheme2}.
Both schemes share a common fundamental structure in which an FP64 matrix is decomposed into multiple low-precision matrices, and the products of these matrices are combined to reconstruct the results of DGEMM.
Because the decomposed low-precision matrices are represented in a fixed-point form that does not involve exponent components, these schemes exhibit particularly high affinity with INT8 MMA units. 
Exploiting this property, several DGEMM emulation methods based on the Ozaki schemes using INT8 MMA units have been proposed~\cite{ootomo2024dgemm,uchino2025Performance,uchino_ozaki2}.

However, the hardware landscape is becoming increasingly heterogeneous. 
While some emerging architectures substantially reduce INT8 computational resources and prioritize low-precision floating-point formats such as FP8, many currently relevant systems still provide comparable FP8 and INT8 throughput, and some accelerators continue to emphasize INT8 support.
Therefore, both paths are practically important: FP8-based emulation is needed to broaden applicability to FP8-oriented architectures, whereas INT8-based emulation remains a strong baseline.
For the Ozaki-I scheme, it is possible to employ FP16 or FP8 MMA units while preserving the same algorithmic structure as that in the INT8-based formulation; implementations exploiting FP16 and FP8 MMA units have been reported~\cite{mukunoki2020,mukunoki2025dgemmfp64arithmetic}. 
In contrast, due to its algorithmic characteristics, the Ozaki-II scheme cannot be directly applied to FP16 or FP8 MMA units with its original algorithmic structure for emulating DGEMM.

\subsection{Our Contributions}
This work enables Ozaki-II-based DGEMM emulation to effectively exploit FP8 MMA units and compares the proposed FP8-based approach with existing INT8-based Ozaki-II emulation.
Our main contributions are:

\paragraph{Explanation of why INT8-based Ozaki-II does not directly translate to FP8}
We review the INT8-based Ozaki-II scheme and identify the algorithmic components that inherently rely on fixed-point and modular semantics, clarifying why a naive replacement of INT8 with FP8 cannot preserve the required exactness.

\paragraph{FP8-based Ozaki-II scheme}
We introduce an FP8-based Ozaki-II scheme that combines (i) a Karatsuba-based extension and (ii) a modular reduction formulation that avoids Karatsuba reconstruction for selected moduli, thereby reducing the number of FP8 matrix multiplications compared to that for a simple Karatsuba-based extension while meeting the target precision.

\paragraph{Rationale for choosing FP8 (not FP16/BF16/FP4)}
We clarify why FP8, rather than FP16, BF16, and FP4, is the appropriate choice for DGEMM emulation by relating the exactness requirement under FP32 accumulation to practical performance considerations.

\paragraph{Comprehensive comparison}
We compare prior FP8-based Ozaki-I and INT8-based Ozaki-II approaches with the proposed method in terms of
(i) effective precision, 
(ii) analytic performance models (including a new model for INT8-based Ozaki-II), 
(iii) working memory footprint (including a new formulation for INT8-based Ozaki-II), and 
(iv) numerical experiments.

Throughout this paper, we assume that INT8 MMA takes INT8 inputs and accumulates in INT32, whereas FP8 MMA takes FP8\_E4M3 inputs and accumulates in FP32.

\section{Ozaki-II Scheme}
The Ozaki-II scheme is a framework for matrix multiplication emulation based on long integer multiplication via the Chinese Remainder Theorem (CRT).
For $N \in \mathbb{N}$, let $p_1, p_2, \dots, p_N \in \mathbb{N}$ be pairwise coprime integers.
The Ozaki-II scheme computes $C \approx AB$ for $A \in \mathbb{R}^{m \times k}$ and $B \in \mathbb{R}^{k \times n}$ via the following steps:
\begin{enumerate}

\item Convert $A$ and $B$ to integer matrices
\begin{align}
    A' &:= \mathrm{trunc}\left(\mathrm{diag}(\mu) \cdot A\right) \in \mathbb{Z}^{m \times k},\label{def:A'}\\
    B' &:= \mathrm{trunc}\left(B \cdot \mathrm{diag}(\nu)\right) \in \mathbb{Z}^{k \times n},\label{def:B'}
\end{align}
where $\mathrm{trunc}(\cdot)$ returns the integer part and $\mu \in \mathbb{R}^m$ and $\nu \in \mathbb{R}^n$ are vectors whose elements are powers of two.
These scaling vectors are determined to satisfy
\begin{equation}\label{condition}
    2\sum_{h=1}^{k} |a'_{ih}||b'_{hj}| < P \quad \forall i,j,
\end{equation}
where $P := \prod_{\ell=1}^N p_\ell$.

\item Compute $C' \approx A'B'$ via the CRT:
\begin{equation}\label{eq:CRT_finalreduction}
    C' \gets \mathrm{mod}\left(\sum_{\ell=1}^N \frac{q_\ell P}{p_\ell} C'_\ell , P \right),
\end{equation}
where $\mathrm{mod}: \mathbb{Z}^{m \times n} \times \mathbb{Z} \to \mathbb{Z}^{m \times n}$ denotes the symmetric modulo operation, $q_\ell$ satisfies $q_\ell P/p_\ell \equiv 1 \pmod{p_\ell}$, and
\begin{equation}\label{eq:CRTmatmul}
    C'_\ell := \mathrm{mod}\left( A'_\ell B'_\ell, p_\ell \right)
\end{equation}
with $A'_\ell := \mathrm{mod}\left(A', p_\ell\right)$ and $B'_\ell := \mathrm{mod}\left(B', p_\ell\right)$.

\item Convert $C'$ back to the floating-point matrix $C$ by applying the following inverse scaling:
\begin{equation}\label{eq:inversescaling}
    C \gets \mathrm{diag}(\mu)^{-1} \cdot C' \cdot \mathrm{diag}(\nu)^{-1}.
\end{equation}

\end{enumerate}

The accuracy of the approximation $C \approx AB$ strongly depends on the relative truncation errors in~\eqref{def:A'} and~\eqref{def:B'} because matrix products in~\eqref{eq:CRTmatmul} can be computed without numerical error.
Therefore, a larger value of $P$ in~\eqref{condition} leads to higher approximation accuracy.
Accordingly, the integers $p_1, p_2, \dots, p_N$ are selected so as to make $P$ as large as possible under practical constraints.
Previous studies have shown that when $p_\ell \le 256$ and $k \le 2^{17}$, the matrix multiplication $A'_\ell B'_\ell$ in~\eqref{eq:CRTmatmul} can be computed using INT8 MMA units~\cite{uchino_ozaki2,uchino_ozaki2_complex}.
In these studies, the moduli $p_\ell$ are selected in descending order from the following set, which is constructed by scanning downward from $256$ and greedily selecting integers that remain pairwise coprime to all previously selected values:
\begin{equation}\label{p_list}
\begin{array}{r@{\ }r@{\ }r@{\ }r@{\ }r@{\ }r@{\ }r@{\ }r@{\ }r@{\ }r@{}l}
\{
 256, &  255, &  253, &  251, &  247, &  241, &  239, &  233, &  229, &  227, \\
 223, &  217, &  211, &  199, &  197, &  193, &  191, &  181, &  179, &  173, \\
 167, &  163, &  157, &  151, &  149, &  139, &  137, &  131, &  127, & \cdots\phantom{,} &  
\}.
\end{array}
\end{equation}
With this selection strategy, $2^7 \le P/2 < 2^{341}$ holds.
By choosing a sufficiently large $N$, the product $P$ becomes large enough to ensure that $A'$ and $B'$ provide sufficiently accurate approximations of $\mathrm{diag}(\mu)\cdot A$ and $B\cdot \mathrm{diag}(\nu)$.
In particular, when $N = 14$, we have $P/2 > 2^{109} > 2^{53 + 53}$, indicating that, in principle, at least $14$ moduli are required to emulate FP64 arithmetic.

\section{Proposed Method Using FP8\_E4M3}

\subsection{Limitations of Direct FP8 Quantization}
Hereafter, we assume that $k \le 2^{16}$ so that no rounding error occurs in FP8\_E4M3 matrix multiplication using FP8 MMA units.
The FP8\_E4M3 format can exactly represent consecutive integers in the range of $-16$ to $16$.
Therefore, following an approach similar to that of the INT8-based Ozaki-II scheme, when $p_\ell \le 32$, $A'_\ell$ and $B'_\ell$ can be represented in FP8\_E4M3 format and the matrix multiplication $A'_\ell B'_\ell$ in~\eqref{eq:CRTmatmul} can be computed using FP8 MMA units.
Possible candidates for moduli satisfying $p_\ell \le 32$ can be selected, for example, from the following set:
\[
\{32,\ 31,\ 29,\ 27,\ 25,\ 23,\ 19,\ 17,\ 13,\ 11,\ 7\}.
\]
Even when all moduli in this set are employed, the resulting product $P$ satisfies $P/2 < 2^{47}$.
As a consequence, the dynamic range of integers reconstructed via the CRT is severely limited, making it difficult to accurately emulate not only FP64 but even FP32 arithmetic.
This fundamental limitation highlights a major challenge in applying the Ozaki-II scheme with FP8\_E4M3 quantization.

\subsection{Karatsuba-Based Extension}
To overcome the limitation, we adopt the Karatsuba method.
Specifically, $A'_\ell$ and $B'_\ell$ are represented as unevaluated sums of two FP8\_E4M3 matrices as follows:
\begin{align*}
A'_\ell = s \cdot A'^{(1)}_\ell + A'^{(2)}_\ell,\quad
B'_\ell = s \cdot B'^{(1)}_\ell + B'^{(2)}_\ell.
\end{align*}
With this decomposition, their product can be expanded as
\begin{equation}\label{prod1}
\begin{aligned}
A'_\ell B'_\ell
&= s^2 A'^{(1)}_\ell B'^{(1)}_\ell\\
&\quad + s\left(A'^{(1)}_\ell B'^{(2)}_\ell + A'^{(2)}_\ell B'^{(1)}_\ell\right)
+ A'^{(2)}_\ell B'^{(2)}_\ell. 
\end{aligned}
\end{equation}
By applying the Karatsuba method, we define $A'^{(3)}_\ell := A'^{(1)}_\ell + A'^{(2)}_\ell$, $B'^{(3)}_\ell := B'^{(1)}_\ell + B'^{(2)}_\ell$, and
\begin{equation}\label{eq:Karatsuba}
C'^{(x)}_\ell := A'^{(x)}_\ell B'^{(x)}_\ell,\quad x \in \{1,2,3\},
\end{equation}
which leads to the following reconstruction:
\begin{equation}
A'_\ell B'_\ell
= s^2 C'^{(1)}_\ell
+ C'^{(2)}_\ell
+ s\left(C'^{(3)}_\ell - C'^{(1)}_\ell - C'^{(2)}_\ell\right).\label{eq:C'-Karatsuba}
\end{equation}
If we have
\begin{equation}\label{limit1}
|(A'_\ell)_{ij}|, |(B'_\ell)_{ij}| \le 256
\end{equation}
and $A'^{(1)}_\ell$ and $B'^{(1)}_\ell$ are obtained as 
\begin{align*}
    (A'^{(1)}_\ell)_{ij} &:= \mathrm{sign}((A'_\ell)_{ij}) \cdot \lceil s^{-1}|(A'_\ell)_{ij}| \rceil,\\
    (B'^{(1)}_\ell)_{ij} &:= \mathrm{sign}((B'_\ell)_{ij}) \cdot \lceil s^{-1}|(B'_\ell)_{ij}| \rceil,
\end{align*}
choosing $s=16$ ensures that $A'^{(x)}_\ell$ and $B'^{(y)}_\ell$ are representable in FP8\_E4M3 for $x,y \in \{1,2,3\}$.
In addition, $|(A'^{(x)}_\ell)_{ij}|, |(B'^{(y)}_\ell)_{ij}| \le 2^4$ holds; we have
\begin{equation}\label{eq:error-free-FP8-matmult}
    \sum_{h=1}^k |(A'^{(x)}_\ell)_{ih}| |(B'^{(y)}_\ell)_{hj}| \le k \cdot 2^4 \cdot 2^4 \le 2^{24}
\end{equation}
for $k \le 2^{16}$. 
From this and $\sum_{h=1}^k (A'^{(x)}_\ell)_{ih} (B'^{(y)}_\ell)_{hj} \in \mathbb{Z}$, all results of arithmetic in the matrix multiplication comply with the definition of FP32; thus, no rounding error occurs in FP8 matrix multiplication in~\eqref{eq:Karatsuba}.
Therefore, the computation of $A'_\ell B'_\ell$ can be realized using three error-free FP8 matrix multiplications.
From~\eqref{limit1}, the coprime integers $p_1, p_2, \dots, p_N$ can be selected to satisfy $p_\ell \le 513$.
Under this setting, the moduli $p_\ell$ can be selected, for example, from the following set, which is constructed by greedily choosing pairwise coprime integers in descending order starting from $513$, and which satisfies $2^8 < P/2 < 2^{713}$:
\begin{equation}\label{p_list_karatsuba}
\begin{array}{r@{\ }r@{\ }r@{\ }r@{\ }r@{\ }r@{\ }r@{\ }r@{\ }r@{\ }r@{}l}
\{
 513, &  512, &  511, &  509, &  505, &  503, &  499, &  493, &  491, &  487, \\
 481, &  479, &  473, &  467, &  463, &  461, &  457, &  449, &  443, &  439, \\
 433, &  431, &  421, &  419, &  409, &  401, &  397, &  389, &  383, &  \cdots\phantom{,} &  
\}.
\end{array}
\end{equation}
For $N \ge 13$, we have $P/2 > 2^{115} > 2^{53+53}$.
Therefore, to achieve precision comparable to that of INT8-based emulation with $14$ moduli, $N \ge 13$ is required in the proposed FP8-based method.

\subsection{Modular Reduction without Karatsuba}
In principle, when two values with 4-bit mantissa are combined, values of up to approximately $2^{4+4+1} = 2^9$ can be represented.
The extra bit comes from allowing the trailing component to be signed: the pair $( (A'^{(1)}_\ell)_{ij}, (A'^{(2)}_\ell)_{ij} )$ forms a redundant signed-digit representation, which can encode roughly one additional bit compared to a naive concatenation of $4+4$ bits.
However, when the Karatsuba method is employed, the sums $A'^{(1)}_\ell + A'^{(2)}_\ell$ and $B'^{(1)}_\ell + B'^{(2)}_\ell$ must be rigorously represented in FP8\_E4M3 format.
This requirement imposes the constraint in~\eqref{limit1}, which restricts the effective representable range.
To relax this limitation, we propose an improved method that exploits modular arithmetic properties without relying on the Karatsuba method.
In~\eqref{prod1}, if $s$ is chosen such that $s^2 = p_\ell$, then we derive $\mathrm{mod}( s^2 ( A'^{(1)}_\ell B'^{(1)}_\ell )_{ij}, p_\ell ) = 0$ because $\mathrm{mod}(s^2, p_\ell) = 0$.
Therefore, for such $s$, we have
\begin{equation}\label{eq:3matmult-notKaratsuba}
\begin{aligned}
C'_\ell 
&= \mathrm{mod}\left( A'_\ell B'_\ell, p_\ell \right)\\
&= \mathrm{mod}\left( s A'^{(1)}_\ell B'^{(2)}_\ell + s A'^{(2)}_\ell B'^{(1)}_\ell
+ A'^{(2)}_\ell B'^{(2)}_\ell, p_\ell \right).
\end{aligned}
\end{equation}
Hence, $C'_\ell$ can be evaluated using three FP8 matrix multiplications without employing the Karatsuba reconstruction.
However, the possible values of $p_\ell = s^2$ are limited.

\subsection{Hybrid Method}\label{subsed:Hybrid_Method}
By combining the proposed modular reduction technique with the Karatsuba method, the moduli $p_\ell$ can be selected, for example, from the following set:
\begin{equation}\label{p_list_hybrid}
\begin{array}{r@{\ }r@{\ }r@{\ }r@{\ }r@{\ }r@{\ }r@{\ }r@{\ }r@{\ }r@{}l}
\{
1089, & 1024, &  961, &  841, &  625, &  529, &  511, &  509, &  503, &  499, \\
 491, &  487, &  481, &  479, &  467, &  463, &  461, &  457, &  449, &  443, \\
 439, &  433, &  431, &  421, &  419, &  409, &  401, &  397, &  389, & \cdots\phantom{,} &  
\}.
\end{array}
\end{equation}
In this construction, we first prioritize square moduli by selecting pairwise coprime squares in descending order from $1089$. We then continue the list with pairwise coprime integers in descending order without restricting them to be squares.
For the square $p_\ell$, we set $s := \sqrt{p_\ell}$. The resulting matrix products are then computed using the proposed modular reduction technique described above.
For these $p_\ell$, $A'^{(1)}_\ell$ and $B'^{(1)}_\ell$ are obtained as 
\begin{align*}
    (A'^{(1)}_\ell)_{ij} &:= \mathrm{round} (s^{-1}(A'_\ell)_{ij}),\\
    (B'^{(1)}_\ell)_{ij} &:= \mathrm{round} (s^{-1}(B'_\ell)_{ij}).
\end{align*}
Then, $A'^{(1)}_\ell$, $A'^{(2)}_\ell$, $B'^{(1)}_\ell$, and $B'^{(2)}_\ell$ are representable in FP8\_E4M3.
In addition, $|(A'^{(x)}_\ell)_{ij}|, |(B'^{(y)}_\ell)_{ij}| \le 2^4$ holds for $x,y \in \{1,2\}$; thus, \eqref{eq:error-free-FP8-matmult} holds for $k \le 2^{16}$ and no rounding error occurs in FP8 matrix multiplication in~\eqref{eq:3matmult-notKaratsuba}.
For the remaining values of $p_\ell$, the Karatsuba-based method is applied.
For this set of moduli, the product $P$ satisfies $2^9 < P/2 < 2^{746}$.
When $N \ge 12$, we obtain $P/2 > 2^{110} > 2^{53+53}$.
Therefore, to achieve precision comparable to that of INT8-based emulation with $N=14$, it is sufficient to choose $N \ge 12$.
Hence, compared to the Karatsuba-based method, the proposed hybrid method can reduce the required number of moduli; the total number of FP8 matrix multiplications is reduced, leading to improved computational efficiency.

\subsection{Method for Conversion to Integer Matrices}
In the first step of the emulation, we convert input floating-point matrices $A \in \mathbb{R}^{m \times k}$ and $B \in \mathbb{R}^{k \times n}$ to integer matrices $A' \in \mathbb{Z}^{m \times k}$ and $B' \in \mathbb{Z}^{k \times n}$, respectively, which satisfy~\eqref{condition}.
In~\cite{uchino_ozaki2,uchino_ozaki2_complex}, two computing modes, namely fast mode and accurate mode, are provided.
Using the scaling vectors $\mu \in \mathbb{R}^{m}$ and $\nu \in \mathbb{R}^n$ in \eqref{def:A'} and \eqref{def:B'}, both modes estimate the upper bound of $2\sum_{h=1}^{k} |a'_{ih}||b'_{hj}|$ in the form
\begin{equation}\label{eq:est_bound}
2 \sum_{h=1}^{k} |a'_{ih}||b'_{hj}| \le 2 \mu_i \bar{w}_{ij} \nu_j,
\end{equation}
where $\bar{w}_{ij}\ge 0$ denotes a mode-dependent computable upper-bound quantity.
Then, the scaling vectors are chosen such that $2 \mu_i \bar{w}_{ij} \nu_j < P$ holds for all $(i,j)$.
In fast mode, the upper bound of $2 \sum_{h=1}^{k} |a'_{ih}||b'_{hj}|$ is obtained based on the Cauchy--Schwarz inequality, even in the FP8-based emulation, where it can be implemented in a way similar to that in the INT8-based emulation.
In accurate mode, the upper bound is estimated by matrix multiplication using INT8 MMA units.
Here, we focus on accurate mode using FP8 matrix multiplication.
Define $\mu' \in \mathbb{R}^{m}$ and $\nu' \in \mathbb{R}^n$ as
\begin{equation}\label{def:mu'nu'}
    \mu'_i := \frac{2^{7}}{\mathrm{ufp}\left( \max_{h}|a_{ih}| \right)},\quad
    \nu'_j := \frac{2^{7}}{\mathrm{ufp}\left( \max_{h}|b_{hj}| \right)},
\end{equation}
where $\mathrm{ufp}(\cdot)$ is the unit in the first place, that is, $\mathrm{ufp}(x) = 2^{\lfloor \log_2 |x| \rfloor}$ for $x \neq 0$.
Note that $\mu'_i$ and $\nu'_j$ are powers of two and $\log_2 \mu'_i$ and $\log_2 \nu'_j$ can be held in the INT16 format.
Then, $\mathrm{diag}(\mu') \cdot A$ and $B \cdot \mathrm{diag}(\nu')$ are cast to the FP8\_E4M3 matrices $\bar{A} \in \mathbb{R}^{m \times k}$ and $\bar{B} \in \mathbb{R}^{k \times n}$, respectively, in round-up mode.
No overflow occurs in the casting because $\mu'_i a_{ij}, \nu'_j b_{ij} < 2^8$.
Let $\bar{C}'$ be the computed result of $\bar{A}\bar{B}$ using FP8 MMA units.
Since each product $\bar{a}_{ih}\bar{b}_{hj}$ is formed in FP32 arithmetic, no rounding error occurs in $\bar{a}_{ih}\bar{b}_{hj}$ for any $i$, $h$, and $j$.
Therefore, rounding errors arise only in the accumulation.
The rounding error in the summation of a floating-point vector $v \in \mathbb{R}^k$ admits the following bound:
\begin{equation}
\left| \sum_{i=1}^k v_i - z \right|
\le (k-1)u \sum_{i=1}^k |v_i|,
\end{equation}
where $z$ is the computed result of the summation and $u$ is the unit roundoff~\cite{rump2012error}.
This implies that
\[
\left|(\bar{A}\bar{B})_{ij}-\bar{c}'_{ij}\right|
\le (k-1)2^{-24}(\bar{A}\bar{B})_{ij}.
\]
That is,
\begin{equation}\label{eq:barCupper}
    \bar{A}\bar{B} \le (1 - (k-1)2^{-24})^{-1} \bar{C}' \le (1 + k2^{-24}) \bar{C}'.
\end{equation}
The factor $(1 - (k-1)2^{-24})^{-1}$ is replaced with the upper bound $(1 + k2^{-24})$ to avoid division.
Therefore, we let $\bar{C}$ be the computed result of $(1 + k2^{-24})\bar{C}'$ in round-up mode. For $\mu \in \mathbb{R}^{m}$ in~\eqref{def:A'} and $\nu \in \mathbb{R}^n$ in~\eqref{def:B'}, we derive
\[
    2\sum_{h=1}^{k} |a'_{ih}||b'_{hj}| 
    \le 2\sum_{h=1}^{k} \mu_i|a_{ih}||b_{hj}|\nu_j
    \le 2\mu_i\mu_i'^{-1} \bar{c}_{ij} \nu_j'^{-1}\nu_j.
\]
Hence, if $\mu$ and $\nu$ are obtained to satisfy
\[
    \mu_i \le \frac{\mu_i'\sqrt{P-1}}{\max_{h}\bar{c}_{ih}},\quad
    \nu_j \le \frac{\nu_j'\sqrt{P-1}}{\max_{h}\bar{c}_{hj}},
\]
$A'$ and $B'$ satisfy~\eqref{condition}.
In the implementation, $\mu$ and $\nu$ are computed as follows:
\begin{align}
    \log_2(\mu_i) &:= \log_2(\mu'_i) + \mathrm{int}(P' + \delta \log_2 \max_{h}\bar{c}_{ih}), \label{eq:mu-computation}\\
    \log_2(\nu_j) &:= \log_2(\nu'_j) + \mathrm{int}(P' + \delta \log_2 \max_{h}\bar{c}_{hj}), \label{eq:nu-computation}
\end{align}
where $P' \in \mathbb{R}$ and $\delta \in \mathbb{R}$ are the FP32 round-down values of
$(\log_2(P-1)-1) / 2$ and $-1/(2-2^{-21})$, respectively, and $P' + \delta \log_2 \max_{h}\bar{c}_{hj}$ is computed using FP32 arithmetic in round-down mode.
The factor $\delta$ compensates for the rounding error of the FP32 $\log_2$ function.

\subsection{Rationale for Choosing FP8 over FP16, BF16, and FP4}

In the present study, we require that no rounding error occurs in FP32 accumulation for low-precision matrix multiplication.
Let $\beta$ denote the number of significand bits of the input format (including the implicit leading bit), and let $k$ be the dot-product length along the $k$-dimension. 
Under the assumption of a common scaling that aligns exponents across products, each term $a_{i\ell}b_{\ell j}$ involves a $\beta$-bit by $\beta$-bit multiplication and therefore needs up to $2\beta$ bits. 
The sum $\sum_{\ell=1}^{k} a_{i\ell}b_{\ell j}$ then requires up to $2\beta+\lceil\log_2 k\rceil$ bits in the worst case. 
Since FP32 provides a 24-bit significand, a necessary condition for rounding-free accumulation in FP32 is
\[
k \cdot 2^{2\beta} \le 2^{24}.
\]
FP8\_E4M3 uses 3 mantissa bits, which corresponds to $\beta=4$ effective significand bits in normal numbers. 
Therefore, for $k \le 2^{16}$, the full FP8\_E4M3 significand can be used while still enabling error-free FP32 accumulation.
In contrast, to keep an FP16/BF16 matrix multiplication error-free at a comparable effective precision (e.g., $\beta \approx 9$), as achieved by the proposed method in Section~\ref{subsed:Hybrid_Method} using three FP8 matrix multiplications, without additional decomposition, the condition above forces $k$-blocking such that each block satisfies $k\le 2^6$. 
However, matrix multiplication throughput degrades significantly for small $k$ 
because the MMA units cannot be fully utilized, reducing arithmetic intensity and making the kernel memory-bound.
When $k$ is large, maintaining exactness requires multiword-style decompositions; in such a regime, an FP16/BF16-based approach can become less efficient than our construction.


This choice is also consistent with hardware trends. 
According to the published specifications for NVIDIA Rubin, the Tensor Core dense specification is 17.5 PFLOP/s for FP8 versus 4.0 PFLOP/s for FP16/BF16 (i.e., a peak ratio of $17.5/4.0 = 4.375$). 
Hence, when the computation can be expressed as FP8 matrix multiplications while preserving the required exactness, relying on FP8 provides a substantially higher peak-throughput envelope than that for FP16/BF16.

Some hardware expose support for FP4 matrix multiplication with FP4\_E2M1 inputs and FP32 accumulation. 
However, FP4 is less natural than FP8 in the present setting. 
Recursive Karatsuba and Toom--Cook require products of sums of low-precision matrices, and under our exactness requirement these intermediate sums must themselves be represented without loss. 
Because FP4 has very limited representational capacity, such intermediate sums cannot in general be maintained exactly within the same FP4 format, so straightforward recursive Karatsuba or Toom--Cook methods are not applicable here. 
By contrast, each FP8 matrix multiplication in the proposed method can in principle be realized by a single-level Karatsuba method using three FP4 matrix multiplications. 
Accordingly, if future hardware provides effective FP4 throughput far exceeding three times that of FP8, such a construction may become advantageous.

\section{Comparison with Previous Work}

\subsection{Number of Low-Precision Matrix Multiplications}
We compare the FP8-based Ozaki-I scheme introduced in~\cite{mukunoki2025dgemmfp64arithmetic}, the proposed FP8-based Ozaki-II scheme, and the INT8-based Ozaki-II scheme introduced in~\cite{uchino_ozaki2,uchino_ozaki2_complex}.
At the time of writing, the absence of a publicly available optimized implementation of the FP8-based Ozaki-I scheme prevents a fair implementation-level comparison. 
Accordingly, we compare against FP8-based Ozaki-I using implementation-agnostic metrics in this paper, and leave a direct empirical comparison as future work when a reproducible implementation is released.

In the FP8-based Ozaki-I scheme, matrices $A$ and $B$ are approximated as unevaluated sums of $S$ instances of FP8 matrices $A_\ell$ and $B_\ell$, respectively, as follows:
\[
A \approx A' := \sum_{\ell=1}^{S} \mathrm{diag}(\zeta^{(\ell)}) A_\ell,\quad
B \approx B' := \sum_{\ell=1}^{S} B_\ell \mathrm{diag}(\eta^{(\ell)}),
\]
where $\zeta^{(\ell)} \in \mathbb{R}^m$ and $\eta^{(\ell)} \in \mathbb{R}^n$ are scaling vectors used to represent $A_\ell$ and $B_\ell$ in the FP8\_E4M3 format.
The matrices $A_\ell$ and $B_\ell$ are obtained such that all matrix products $A_i B_j$ are computed using FP8 MMA units without rounding errors.
The approximation of $AB$ is then obtained as
\begin{equation}\label{eq:Ozaki-I-all}
\sum_{i=1}^{S}\sum_{j=1}^{S}\mathrm{diag}(\zeta^{(i)}) \cdot A_i B_j \cdot \mathrm{diag}(\eta^{(j)}).
\end{equation}
Each slice represents 4 bits of effective information and an extra bit exists between adjacent slices.
Therefore, the total effective precision is given by $4S + (S-1) = 5S-1$ bits.
To emulate FP64 arithmetic, $5S - 1 \ge 53$ should be satisfied, which requires at least 11 slices.
Consequently, the minimum number of FP8 matrix multiplications in the Ozaki-I scheme is $11^2 = 121$. In the INT8-based Ozaki-I scheme~\cite{ootomo2024dgemm,uchino2025Performance}, instead of evaluating~\eqref{eq:Ozaki-I-all}, the computation is performed by neglecting smaller terms:
\[
\sum_{i+j \le S+1}\mathrm{diag}(\zeta^{(i)}) \cdot A_i B_j \cdot \mathrm{diag}(\eta^{(j)}).
\]
This modification reduces the number of matrix multiplications to $S(S+1)/2$; however, it introduces a loss of accuracy. 
We refer to the methods involving $S(S+1)/2$ and $S^2$ matrix multiplications as fast mode and accurate mode, respectively.
In contrast, the proposed method achieves 53 or more bits of precision with $N \ge 12$, requiring only $12 \times 3 = 36$ FP8 matrix multiplications.
Note that in accurate mode, one additional FP8 matrix multiplication is required.
In the INT8-based Ozaki-II scheme, $N \ge 14$ is required to achieve 53 or more bits of precision.
The number of moduli is greater than that of the proposed FP8-based method. 
The number of INT8 matrix multiplications is $N$ in fast mode and $N+1$ in accurate mode.

Table~\ref{tab:comparison_ozaki_fp8} summarizes the effective precision and the number of FP8/INT8 matrix multiplications required for these methods.
Compared to the FP8-based Ozaki-I scheme, the proposed approach achieves comparable or higher precision while significantly reducing the number of multiplications.
However, compared to the INT8-based Ozaki-II scheme, the proposed approach requires approximately $2.5 \times$ the number of matrix multiplications.

\begin{table}[htbp]
\centering
\caption{Comparison of number of FP8/INT8 matrix multiplications and effective precision of fixed-point matrices $A'$ and $B'$\label{tab:comparison_ozaki_fp8}}
\begin{tabular}{cccc}
\toprule
\multirow{2}{*}{Method} & \multicolumn{2}{c}{\#MatMults} & \multirow{2}{*}{Effective Bits} \\
& fast & accurate & \\
\midrule
FP8 Ozaki-I ($S$ slices) & $S(S+1)/2$ & $S^2$ & $\lesssim 5S-1$ \\
FP8 Ozaki-I ($11$ slices) & $66$ & $121$ & $\lesssim 54$ \\
FP8 Ozaki-I ($12$ slices) & $78$ & $144$ & $\lesssim 59$ \\
FP8 Ozaki-I ($13$ slices) & $91$ & $169$ & $\lesssim 64$ \\
\midrule
FP8 Ozaki-II ($N$ moduli) & $3N$ & $3N+1$ & $\lesssim \log_2 \sqrt{P/2}$ \\
FP8 Ozaki-II ($12$ moduli) & $36$ & $37$ & $\lesssim 55$ \\
FP8 Ozaki-II ($13$ moduli) & $39$ & $40$ & $\lesssim 59$ \\
FP8 Ozaki-II ($14$ moduli) & $42$ & $43$ & $\lesssim 64$ \\
\midrule
INT8 Ozaki-II ($N$ moduli) & $N$ & $N+1$ & $\lesssim \log_2 \sqrt{P/2}$ \\
INT8 Ozaki-II ($14$ moduli) & $14$ & $15$ & $\lesssim 54$ \\
INT8 Ozaki-II ($15$ moduli) & $15$ & $16$ & $\lesssim 58$ \\
INT8 Ozaki-II ($16$ moduli) & $16$ & $17$ & $\lesssim 62$ \\
\bottomrule
\end{tabular}
\end{table}

\subsection{Performance Modeling and Throughput Comparison}
\label{subsec:Performance Modeling and Throughput Comparison}
Next, we construct performance models for the INT8-based Ozaki-II scheme and for the proposed FP8-based method based on their available implementations, and compare their predicted throughput. While a performance model is provided for the complex-valued Ozaki-II extension, the corresponding real-valued INT8-based Ozaki-II studies primarily report empirical results and do not present an explicit analytic performance model; we address this gap by developing such a model for real-valued general matrix--matrix multiplication (GEMM) emulation.

Let $b$ denote the sustained memory bandwidth in bytes/s.
Let $\mathrm{OPS}_{\mathrm{i8}}$ and $\mathrm{OPS}_{\mathrm{f8}}$ denote the sustained compute throughput in (FL)OP/s of INT8 and FP8 GEMM using the BLAS (Basic Linear Algebra Subprograms) implementation, respectively.
We treat $c$ as a platform-dependent, dimensionless correction parameter that summarizes FP64- and control-heavy overheads not captured by the GEMM and bandwidth terms (e.g., modular arithmetic and CRT reconstruction). 
For example, $c$ is set on the order of the number of low-precision matrix multiplications, reflecting that many overhead components are invoked once per modulus. 
Following the modeling approach used for the complex-valued INT8-based Ozaki-II scheme, we derive the time model $T_{\mathrm{i8fast}}$ for the INT8-based Ozaki-II scheme in fast mode as follows:
\begin{align*}
T_{\mathrm{i8fast}}(m,n,k,N,c)
&= \frac{2mnkN}{\mathrm{OPS}_{\mathrm{i8}}}
+ \frac{(12 + 6N + 2c)mn}{b}\\
&\quad + \frac{((16 + N + c)k + 2)(m + n)}{b}.
\end{align*}
Similarly, we obtain the time model $T_{\mathrm{i8acc}}$ for accurate mode as follows:
\begin{align*}
&T_{\mathrm{i8acc}}(m,n,k,N,c)\\
&= \frac{2mnk(N + 1)}{\mathrm{OPS}_{\mathrm{i8}}}
+ \frac{(20 + 6N + 2c)mn}{b}\\
&\quad + \frac{((17 + N + c)k + 4)(m + n) + 2km + 2n}{b}.
\end{align*}

Following the algorithmic flow of the proposed FP8-based Ozaki-II scheme, we first derive the time model for accurate mode, denoted by $T_{\mathrm{f8acc}}$.
The corresponding fast-mode model $T_{\mathrm{f8fast}}$ is obtained analogously by omitting the additional procedure; therefore, we present only the final expression.
\begin{enumerate}
    \item We load the input matrices $A \in \mathbb{R}^{m \times k}$ and $B \in \mathbb{R}^{k \times n}$ once to compute $\mu'$ and $\nu'$ defined in~\eqref{def:mu'nu'} and to form the FP8\_E4M3 matrices $\bar{A} \in \mathbb{R}^{m \times k}$ and $\bar{B} \in \mathbb{R}^{k \times n}$.
    We then store the base-2 exponents of $\mu'$ and $\nu'$ as INT16 arrays, along with the FP8\_E4M3 matrices $\bar{A}$ and $\bar{B}$.

    \item We compute the FP8 matrix product $\bar{C}' := \bar{A}\bar{B}$.

    \item\label{item:3} We load $\bar{C}'$ and $\mu'$ and compute the scaling vector $\mu \in \mathbb{R}^m$ as in~\eqref{eq:mu-computation}, and then store $\mu$.
    
    \item\label{item:4} We load $A$ and $\mu$.
    We then convert $A$ to $A'$ as in~\eqref{def:A'} and compute the modular residues $A'_\ell$ for $N$ moduli.

    \item\label{item:5} For each square modulus, we store two FP8\_E4M3 matrices $A'^{(1)}_\ell, A'^{(2)}_\ell \in \mathbb{Z}^{m \times k}$ and for each non-square modulus, we store three FP8\_E4M3 matrices $A'^{(1)}_\ell, A'^{(2)}_\ell, A'^{(3)}_\ell \in \mathbb{Z}^{m \times k}$.
    
    \item We apply a similar procedure as in Steps~\ref{item:3}--\ref{item:5} to compute $\nu \in \mathbb{R}^n$ and to obtain $B'_\ell$ and store the corresponding FP8\_E4M3 matrices.
    
    \item For each square modulus, we compute the FP8 matrix products $C'^{(1)}_\ell := A'^{(1)}_\ell B'^{(2)}_\ell$, $C'^{(2)}_\ell := A'^{(2)}_\ell B'^{(1)}_\ell$, and $C'^{(3)}_\ell := A'^{(2)}_\ell B'^{(2)}_\ell$ without rounding error. We then compute $C'_\ell$ as in~\eqref{eq:3matmult-notKaratsuba} and store it as an INT16 matrix.
    
    \item For each nonsquare modulus, we compute the FP8 matrix products as in~\eqref{eq:Karatsuba} without rounding error. We then compute $C'_\ell$ based on~\eqref{eq:C'-Karatsuba} and store it as an INT16 matrix.

    \item We load $C'_\ell$ for all $\ell=1,\dots,N$. We then compute $C' \approx A'B'$ as in~\eqref{eq:CRT_finalreduction}.

    \item We load $\mu$ and $\nu$. We then compute $C \approx AB$ as in~\eqref{eq:inversescaling} and store it.
\end{enumerate}
We assume that $N < 34$, that is, the square moduli are $p_1,p_2,\dots,p_6$.
Based on the workflow, for 
\begin{equation}\label{eq:M}
    M_N := 
    \begin{cases}
        2N & (\text{if $N \le 6$}),\\
        3N-6 & (\text{otherwise})
    \end{cases}
\end{equation}
indicating the number of $A'^{(x)}_\ell$ or $B'^{(x)}_\ell$, we derive 
\begin{align*}
&T_{\mathrm{f8acc}}(m,n,k,N,c)\\
&= \frac{2mnk(N + 1)}{\mathrm{OPS}_{\mathrm{f8}}}
+ \frac{(20 + 2c + 4N + 4M_N)mn}{b}\\
&\quad + \frac{((17 + M_N + c)k + 4)(m+n) + 2km + 2n}{b}.
\end{align*}
Similarly, we obtain $T_{\mathrm{f8fast}}$ as
\begin{align*}
&T_{\mathrm{f8fast}}(m,n,k,N,c)\\
&= \frac{2mnkN}{\mathrm{OPS}_{\mathrm{f8}}}
+ \frac{(12 + 2c + 4N + 4M_N)mn}{b}\\
&\quad + \frac{((16 + M_N + c)k + 2)(m + n)}{b}.
\end{align*}

\begin{figure}[htb]\centering
\noindent
\begin{minipage}[b]{.49\hsize}
\includegraphics[width=\hsize]{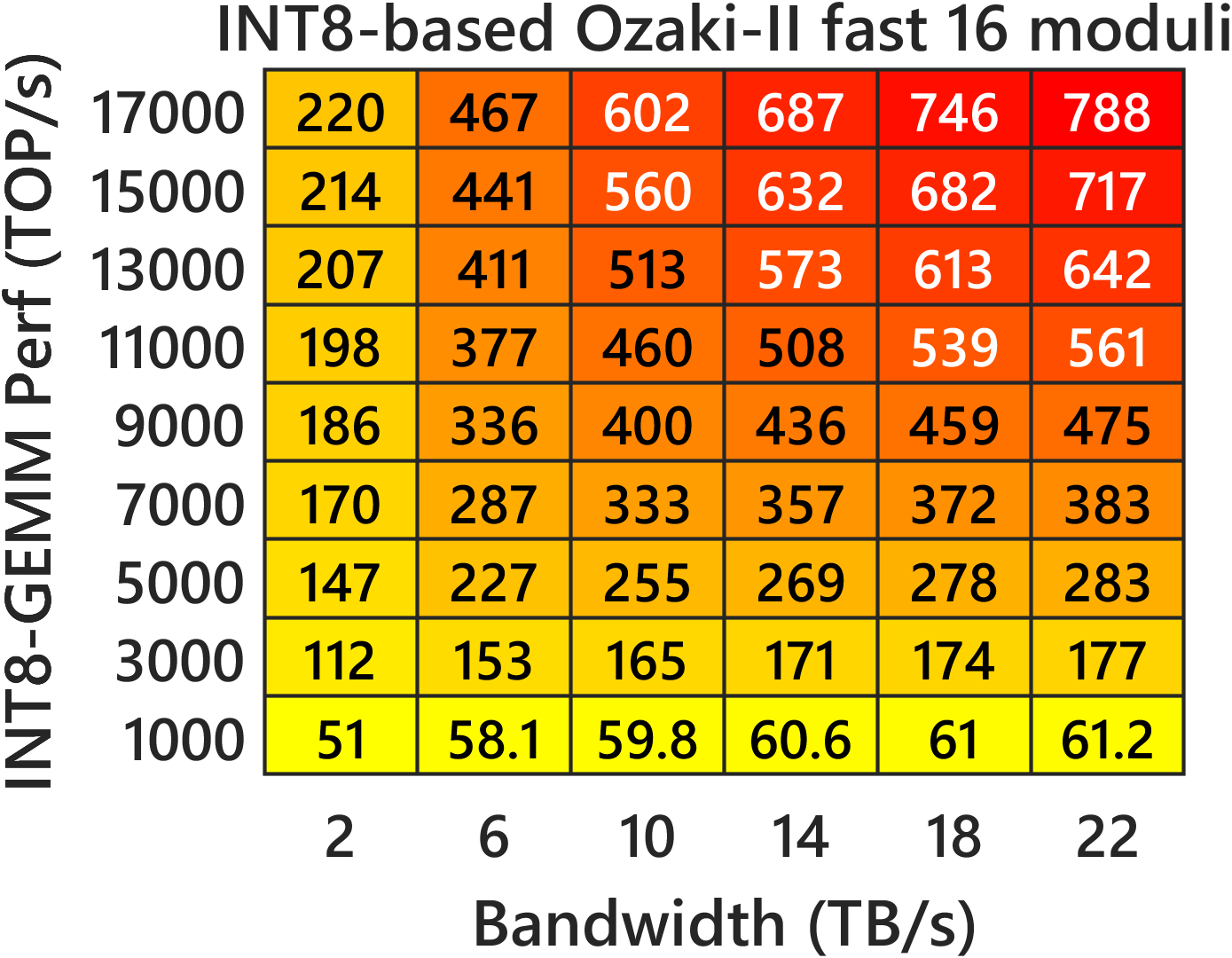}
\end{minipage}
\begin{minipage}[b]{.49\hsize}
\includegraphics[width=\hsize]{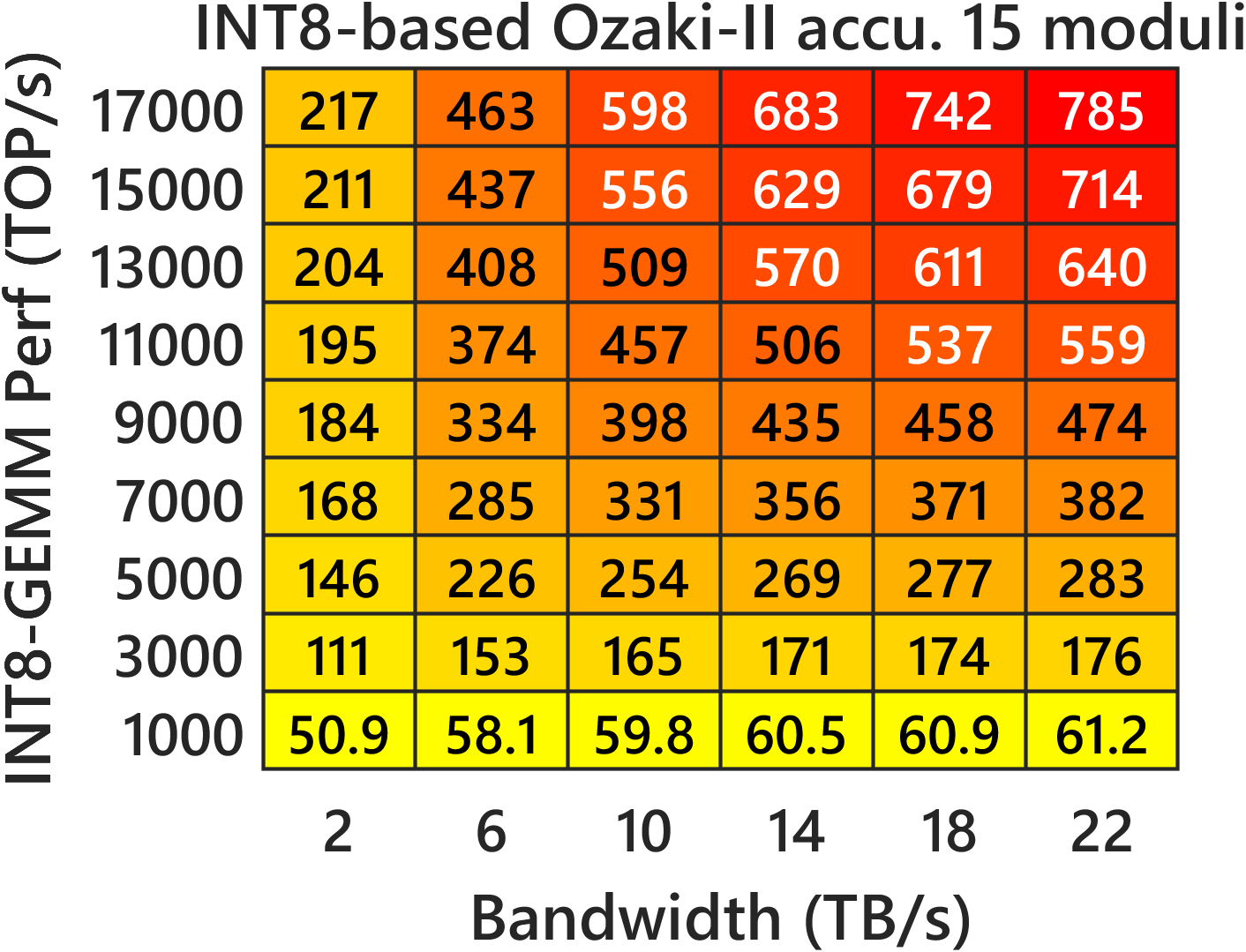}
\end{minipage}
\caption{Predicted throughput heatmaps for DGEMM emulation using INT8-based Ozaki-II scheme, shown as functions of sustained BLAS GEMM throughput and sustained memory bandwidth. Left: fast mode ($T_{\mathrm{i8fast}}(16384,16384,16384,16,16)$). Right: accurate mode ($T_{\mathrm{i8acc}}(16384,16384,16384,15,16)$). The correction terms are equal to the numbers of matrix multiplications.}
\label{fig:heatmap_i8_model}
\end{figure}

\begin{figure}[htb]\centering
\noindent
\begin{minipage}[b]{.49\hsize}
\includegraphics[width=\hsize]{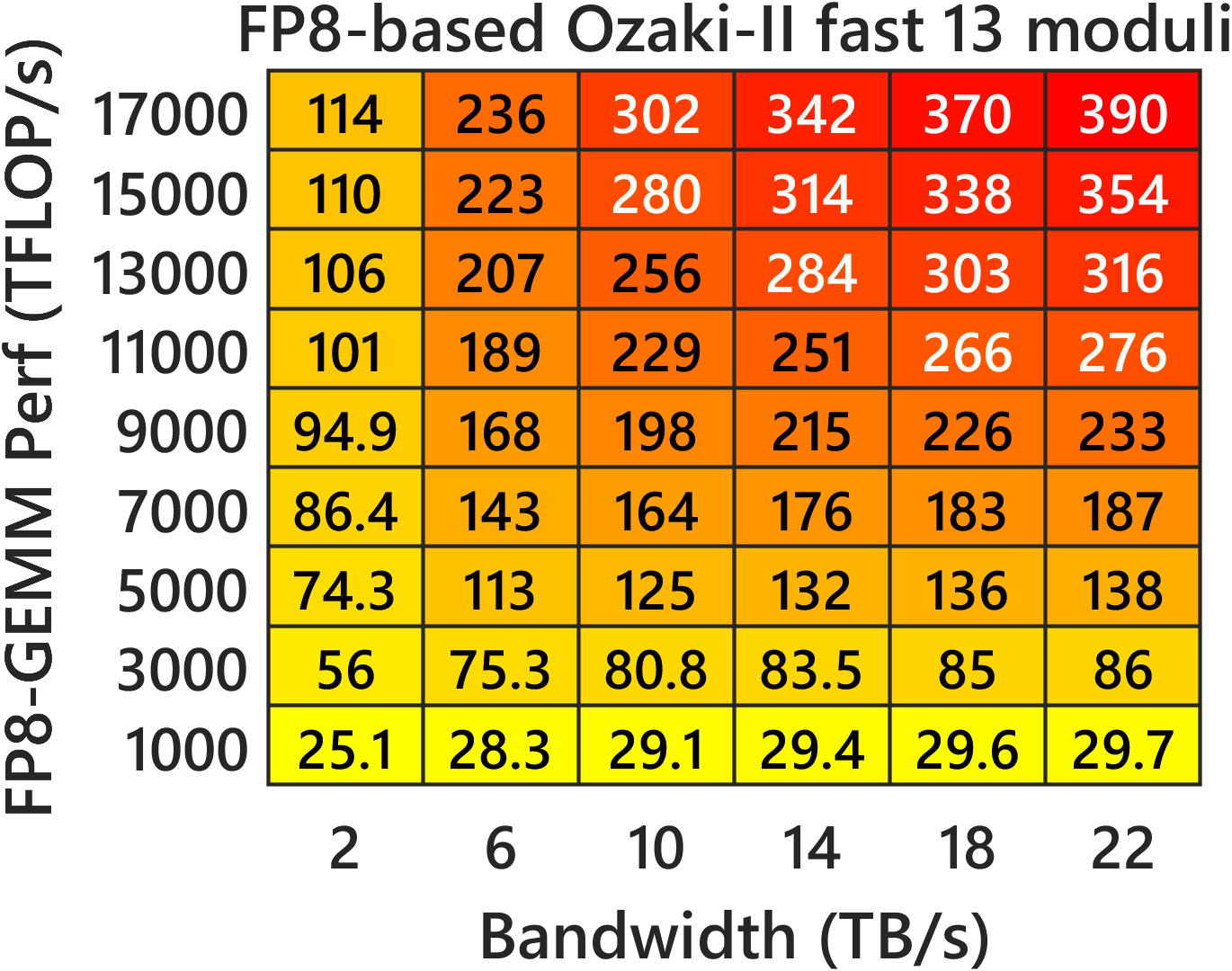}
\end{minipage}
\begin{minipage}[b]{.49\hsize}
\includegraphics[width=\hsize]{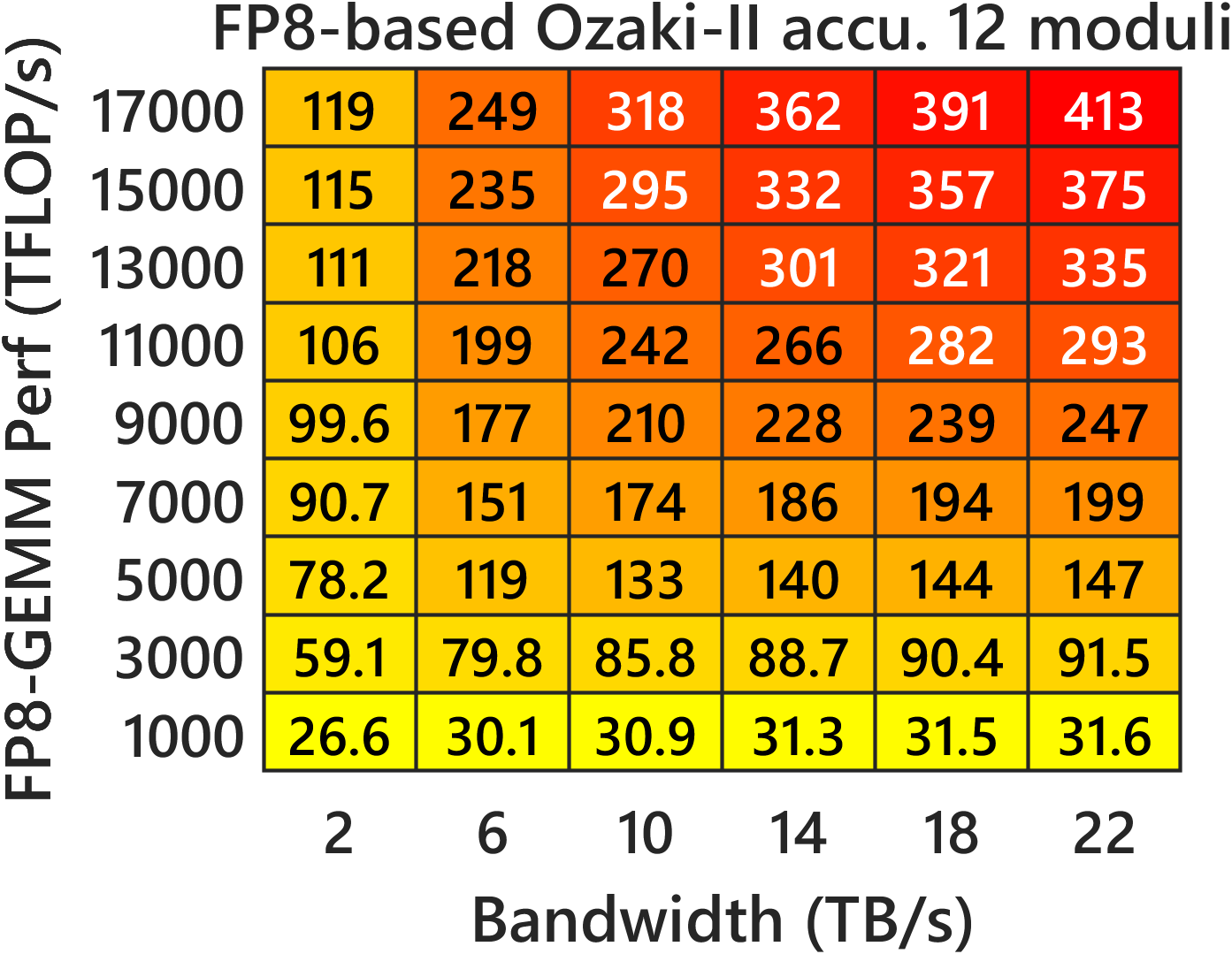}
\end{minipage}
\caption{Predicted throughput heatmaps for DGEMM emulation using proposed FP8-based Ozaki-II scheme, shown as functions of sustained BLAS GEMM throughput and sustained memory bandwidth. Left: fast mode ($T_{\mathrm{f8fast}}(16384,16384,16384,13,39)$). Right: accurate mode ($T_{\mathrm{f8acc}}(16384,16384,16384,12,37)$). The correction terms are equal to the numbers of matrix multiplications.}
\label{fig:heatmap_f8_model}
\end{figure}

For reference, Figs.~\ref{fig:heatmap_i8_model} and~\ref{fig:heatmap_f8_model} present heatmaps of the predicted throughput obtained from the above models for the INT8-based and the proposed FP8-based Ozaki-II schemes, respectively.
NVIDIA's published Rubin GPU specification reports that the peak performance of FP64 DGEMM using emulation is 200~TFLOP/s~\cite{Rubin}. 
The performance prediction in Fig.~\ref{fig:heatmap_f8_model} suggests that, under Rubin-like hardware specifications, the proposed FP8-based Ozaki-II emulation could exceed this 200~TFLOP/s reference level by a substantial margin.
In addition, the predicted throughput indicates that, if the throughput of the FP8 matrix multiplication is only about a factor of two faster than that of the INT8 matrix multiplication, the INT8-based emulation will likely remain faster.

\subsection{Working Memory Footprint Comparison}
Next, we compare the working memory footprint of the INT8-based Ozaki-II scheme and the proposed FP8-based method.
Here, we use \emph{workspace} to refer to the additional temporary buffers required during the computation, excluding the input and output matrices, and \emph{working memory footprint} to refer to the total size of this workspace.

In the INT8-based Ozaki-II scheme with $N$ moduli, for the input matrix $A \in \mathbb{R}^{m \times k}$, the method stores $N$ INT8 residue matrices of size $m \times k$ and one INT16 scaling vector of size $m$.
Similarly, for the input matrix $B \in \mathbb{R}^{k \times n}$, $N$ INT8 residue matrices of size $k \times n$ and one INT16 scaling vector of size $n$ are required.
For the matrix multiplication stage, the matrix products are stored in $N$ INT32 matrices of size $m \times n$.
$N$ INT8 matrices of size $m \times n$ for the reduced residues after modular reduction are also stored.
Therefore, the working memory footprint in bytes is
\begin{equation}\label{eq:W8i}
W_{\mathrm{i8}}(m,n,k,N) = (mk + kn + 5mn)N + 2(m+n).
\end{equation}

In the FP8-based Ozaki-II scheme with $N$ moduli, with $M$ defined in~\eqref{eq:M},  for the input matrix $A$, the method stores $M$ FP8 residue matrices of size $m \times k$ and one INT16 scaling vector of size $m$.
Similarly, for the input matrix $B$, $M$ FP8 residue matrices of size $k \times n$ and one INT16 scaling vector of size $n$ are required.
For the matrix multiplication stage, the matrix products are stored in $M$ INT32 matrices of size $m \times n$.
$N$ INT16 matrices of size $m \times n$ for the reduced residues after modular reduction are also stored.
Therefore, the working memory footprint in bytes is
\begin{equation}\label{eq:W8f}
\begin{aligned}
&W_{\mathrm{f8}}(m,n,k,N)\\
&= (mk + kn + 4mn)\cdot M_N + 2Nmn + 2(m + n).
\end{aligned}
\end{equation}

We note that in practice, all matrix dimensions are padded to multiples of 256, both to avoid implementation-specific issues in the native BLAS INT8/FP8 matrix multiplication kernels and to ensure high performance for the overall workload.
A comparison of the two models shows that the FP8-based method requires a larger working memory footprint than that for the INT8-based Ozaki-II scheme.
This increase is mainly due to the additional temporary buffers needed to represent each residue using multiple FP8 matrices and to store intermediate results in INT16 instead of INT8.
In particular, the former overhead arises because each residue is inherently a fixed-point quantity that does not require an exponent field, whereas storing it in the FP8 format leaves the exponent bits effectively unused.
In large-scale HPC applications, the workload consumes substantial workspace for multiple kernels and auxiliary data structures, leaving limited memory headroom for additional temporary buffers.
From~\eqref{eq:W8i} and~\eqref{eq:W8f}, both emulation methods require a large working memory.
For example, with $m=n=k=16384$, the INT8-based Ozaki-II scheme with $N=14$ requires 27 GB and the FP8-based Ozaki-II scheme with $N=12$ requires 55 GB.
From this perspective, reducing the workspace overhead of DGEMM emulation is an important practical challenge.

For large problem sizes $(m,n,k)$, reducing the working memory footprint can be achieved for both the INT8- and FP8-based methods by blocking the emulation call.
As a first-order estimate, if the emulation is executed on blocked subproblems of size $(m_{\mathrm{blk}}, n_{\mathrm{blk}}, k_{\mathrm{blk}})$, the total execution time in fast mode can be approximated as
\[
T_{\mathrm{i8fast}}(m_{\mathrm{blk}}, n_{\mathrm{blk}}, k_{\mathrm{blk}}, N, c)
\times \left\lceil \frac{m}{m_{\mathrm{blk}}} \right\rceil
\times \left\lceil \frac{n}{n_{\mathrm{blk}}} \right\rceil
\times \left\lceil \frac{k}{k_{\mathrm{blk}}} \right\rceil.
\]
The same approximation applies to the other models ($T_{\mathrm{i8acc}}$, $T_{\mathrm{f8fast}}$, and $T_{\mathrm{f8acc}}$).
Here, edge effects and additional accumulation overhead across $k$-blocks are omitted for simplicity.
However, because native INT8/FP8 matrix multiplication kernels typically do not achieve high throughput for small matrices, blocking all of $m$, $n$, and $k$ is not desirable in practice.
In particular, reducing $k$ tends to lower arithmetic intensity and underutilize the MMA units, which can significantly degrade throughput.
From~\eqref{eq:W8i} and~\eqref{eq:W8f}, blocking only $m$ and $n$, while keeping $k$ large, can still effectively reduce the working memory footprint, while suppressing performance degradation.
Therefore, blocking in the $m$ and $n$ dimensions is a practical and effective strategy for workspace reduction in both schemes.

\section{Numerical Results}
\label{sec:Numerical Results}
The proposed method is provided as an open-source library\footnote{%
\ifanonymous%
URL omitted due to double-anonymous review; it will be included in the camera-ready version. %
\else%
https://github.com/RIKEN-RCCS/GEMMul8%
\fi%
}.
The library supports both the INT8-based Ozaki-II scheme and the proposed FP8-based Ozaki-II scheme, and yields bitwise reproducible results under a fixed toolchain. 
All experiments in this section were conducted using this library.

All experiments were conducted on the following platforms. 
\begin{itemize}
    \item An NVIDIA GeForce RTX~4090 Laptop GPU with an Intel Core i9-14900HX CPU, using CUDA Toolkit 13.2.51.
    \item An NVIDIA GeForce RTX~5080 GPU with an AMD Ryzen 9 7950X CPU, using CUDA Toolkit 13.2.51.
    \item An AMD Radeon RX 9070 XT GPU with an Intel Core i7-7820X CPU, using ROCm 7.2.1.
    \item An NVIDIA GH200 Grace Hopper Superchip with CUDA Toolkit 13.1.115.
    \item An NVIDIA GB10 Grace Blackwell Superchip with CUDA Toolkit 13.2.51.
    \item An NVIDIA B200 SXM GPU with an Intel Xeon 6960P CPU in SAKURA internet Inc.'s managed HPC cluster service, SAKURAONE~\cite{SAKURAONE}, using CUDA Toolkit 13.1.80.
\end{itemize}
For throughput and time-breakdown measurements, we performed 30 warm-up runs followed by 30 timed runs for each configuration. The median is reported.
Native FP64 DGEMM was computed using \texttt{cublasDgemm} implemented in cuBLAS. 
The underlying FP8 matrix multiplications inside the emulation were executed using cuBLASLt, whereas the INT8 matrix multiplications were executed using cuBLAS.

\subsection{Accuracy}
\begin{figure}[htbp]
\centering
\includegraphics[width=\hsize]{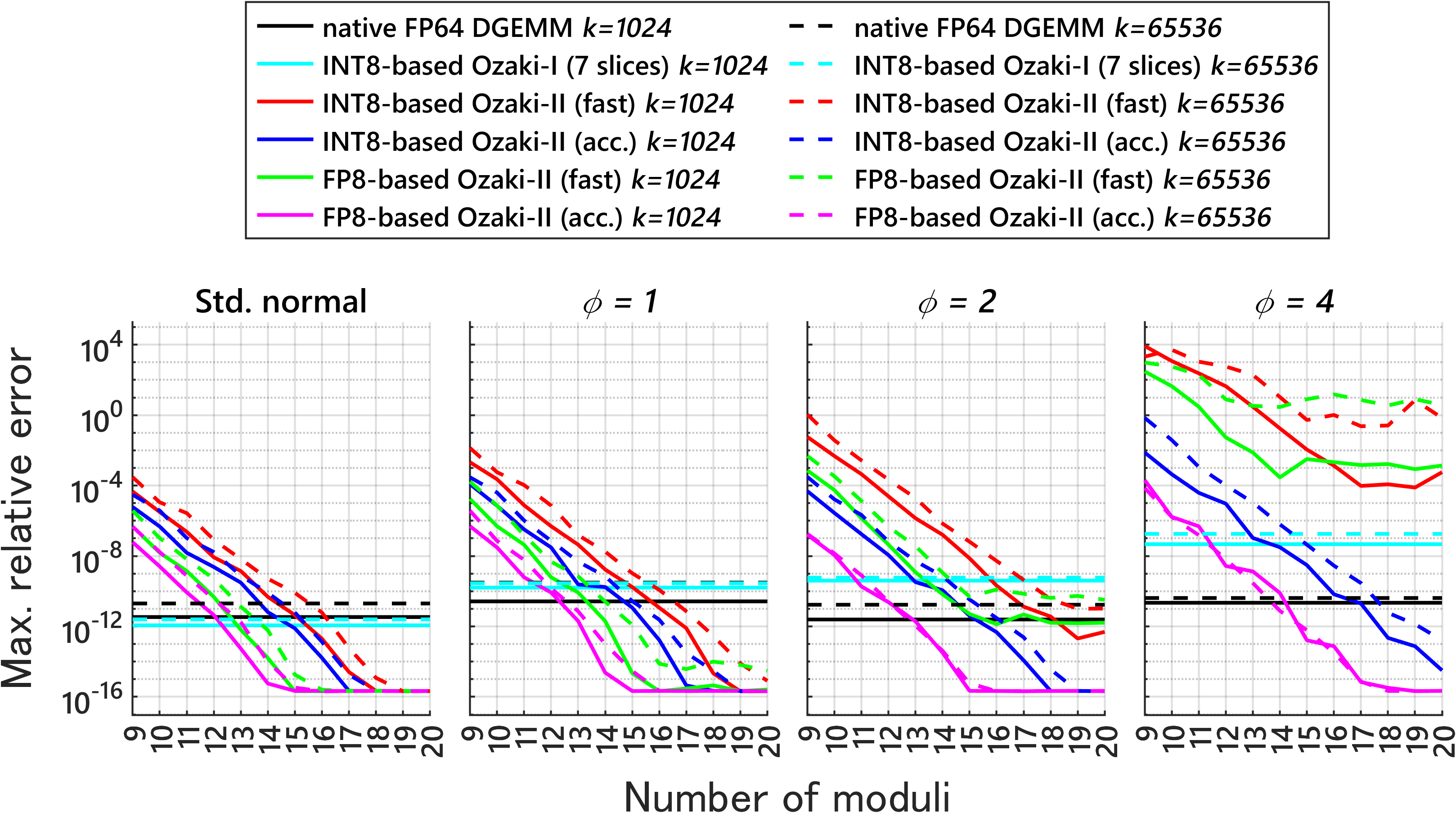}
\caption{Accuracy comparison on RTX 4090 Laptop.\label{fig:accuracy}}
\end{figure}

Fig.~\ref{fig:accuracy} shows the accuracy of DGEMM on the RTX 4090 Laptop.
Similar results were obtained on other platforms.
For the INT8-based Ozaki-I baseline, we used the cuBLAS implementation.
To control the dynamic range of the test matrices, we generated $A \in \mathbb{R}^{128 \times k}$ and $B \in \mathbb{R}^{k \times 128}$ as $a_{ij}, b_{ij} \approx (\mathrm{rand}-0.5)\cdot \exp(\mathrm{randn}\cdot \phi)$, where $\mathrm{rand}\in(0,1]$ denotes a uniformly distributed random number, $\mathrm{randn}$ denotes a standard normal random number, and $\phi$ controls the spread of magnitudes. 
For the plot labeled ``Std. normal'', the matrix entries were generated as standard normal random numbers.
All random numbers were generated using the cuRAND API.

In fast mode, the upper bound in~\eqref{eq:est_bound} is estimated using the Cauchy--Schwarz inequality, which can be overly conservative. 
In contrast, in accurate mode, the bound is estimated using a direct low-precision matrix multiplication (INT8 or FP8), enabling the scaling vectors to be chosen such that the estimated bound is closer to $P/2$. 
As a result, accurate mode typically yields higher accuracy. 

When the test matrices have a narrower dynamic range, such as normally distributed matrices or matrices generated with small $\phi$, the overestimation in fast mode is smaller, and the accuracy gap between the two modes correspondingly decreases. 
For reference, it has been reported that, for DGEMM in many real-world applications, the INT8-based Ozaki-I scheme using unsigned slice encoding~\cite{NVIDIA2026DGEMM}, as implemented in cuBLAS, is sufficient with seven 8-bit slices~\cite{GTC2025NVIDIA}, corresponding to approximately $55$ effective bits.
For test matrices with normally distributed random entries, the INT8-based Ozaki-II emulation with 15 moduli in fast mode and 16 moduli in accurate mode and the FP8-based Ozaki-II emulation with 13 moduli in fast mode and 12 moduli in accurate mode achieve accuracy close to that of the cuBLAS Ozaki-I baseline with seven slices.

In accurate mode, the accuracy gap between $k=1024$ and $k=65536$ was smaller for the proposed FP8-based emulation than for the INT8-based emulation. 
We attribute this to the bound estimation step. 
In the INT8-based emulation, the inputs are converted to INT8 matrices using a ceiling operation as $\lceil \mathrm{diag}(\mu')A\rceil$ and $\lceil B\,\mathrm{diag}(\nu')\rceil$ to avoid underestimation; this can inflate small-magnitude entries to $1$, leading to a looser bound in~\eqref{eq:est_bound}. 
In contrast, the proposed method converts the inputs to FP8\_E4M3 in round-up mode, which introduces smaller inflation for small values and thus enables a tighter evaluation of the bound.

\subsection{Throughput}
\begin{figure}[htbp]
\centering
\includegraphics[width=\hsize]{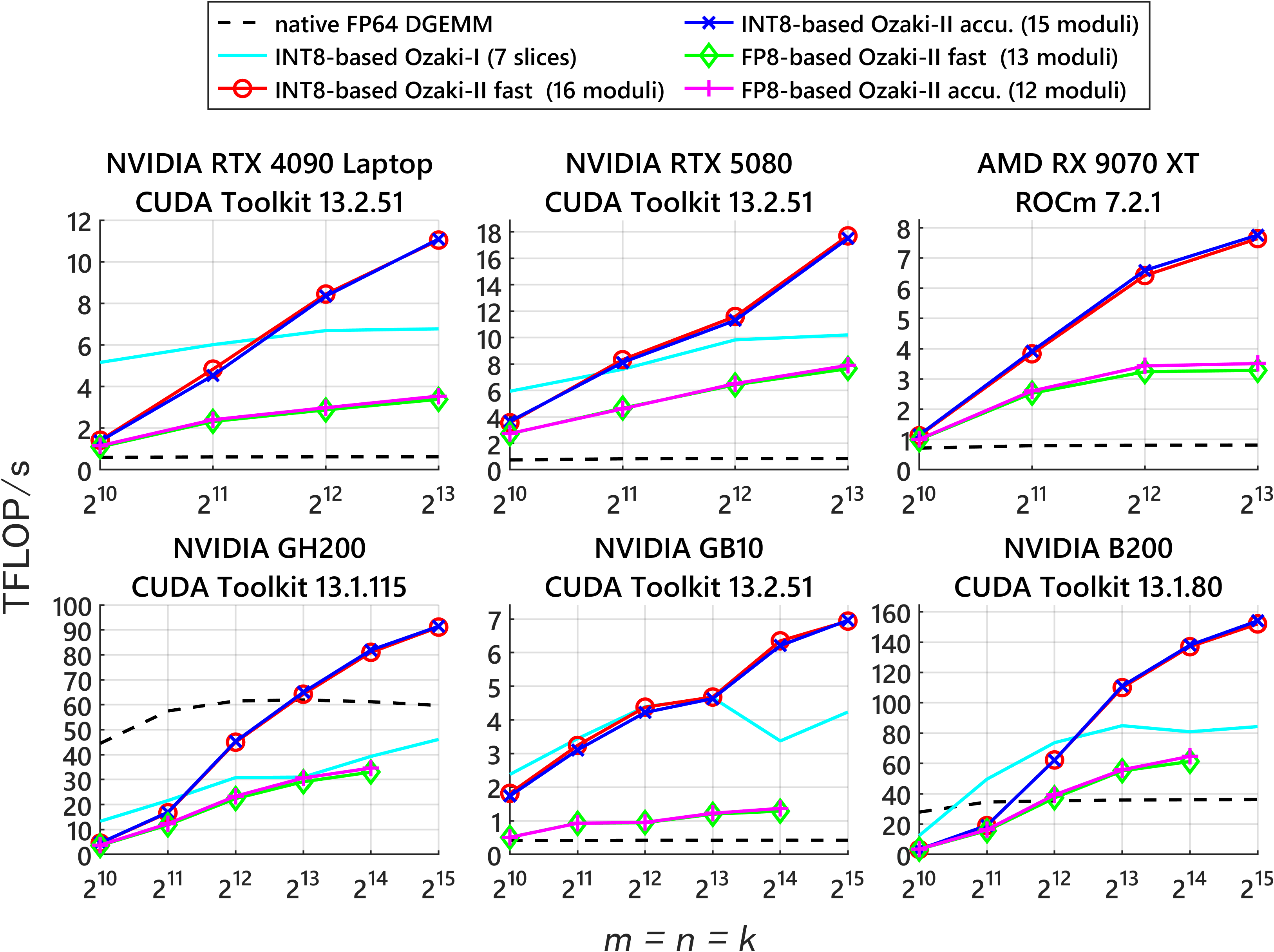}
\caption{Throughput comparison on multiple platforms.\label{fig:flops_square_multi}}
\end{figure}

We first compare DGEMM throughput for square matrices across a diverse set of platforms. 
Fig.~\ref{fig:flops_square_multi} provides a cross-platform overview for $m=n=k$, allowing the effect of hardware characteristics on the relative performance of native FP64 DGEMM and the emulation schemes.
A clear platform dependence is observed. 
On low-FP64-throughput GPUs, such as the RTX~4090~Laptop, RTX~5080, RX~9070~XT, and GB10, all emulation schemes provide substantially higher throughput than native FP64 DGEMM even for $m=n=k=1024$. 
In contrast, on platforms with higher native FP64 capability, such as the GH200 and B200, the performance advantage of emulation becomes smaller and is limited to more favorable parameter ranges.
For sufficiently large problems, the following order of speed was observed across all environments except for GH200: INT8-based Ozaki-II, INT8-based Ozaki-I, FP8-based Ozaki-II, and native FP64.
This exception on the GH200 is attributable to its relatively strong native FP64 capability.

\begin{figure}[htbp]
\centering
\includegraphics[width=\hsize]{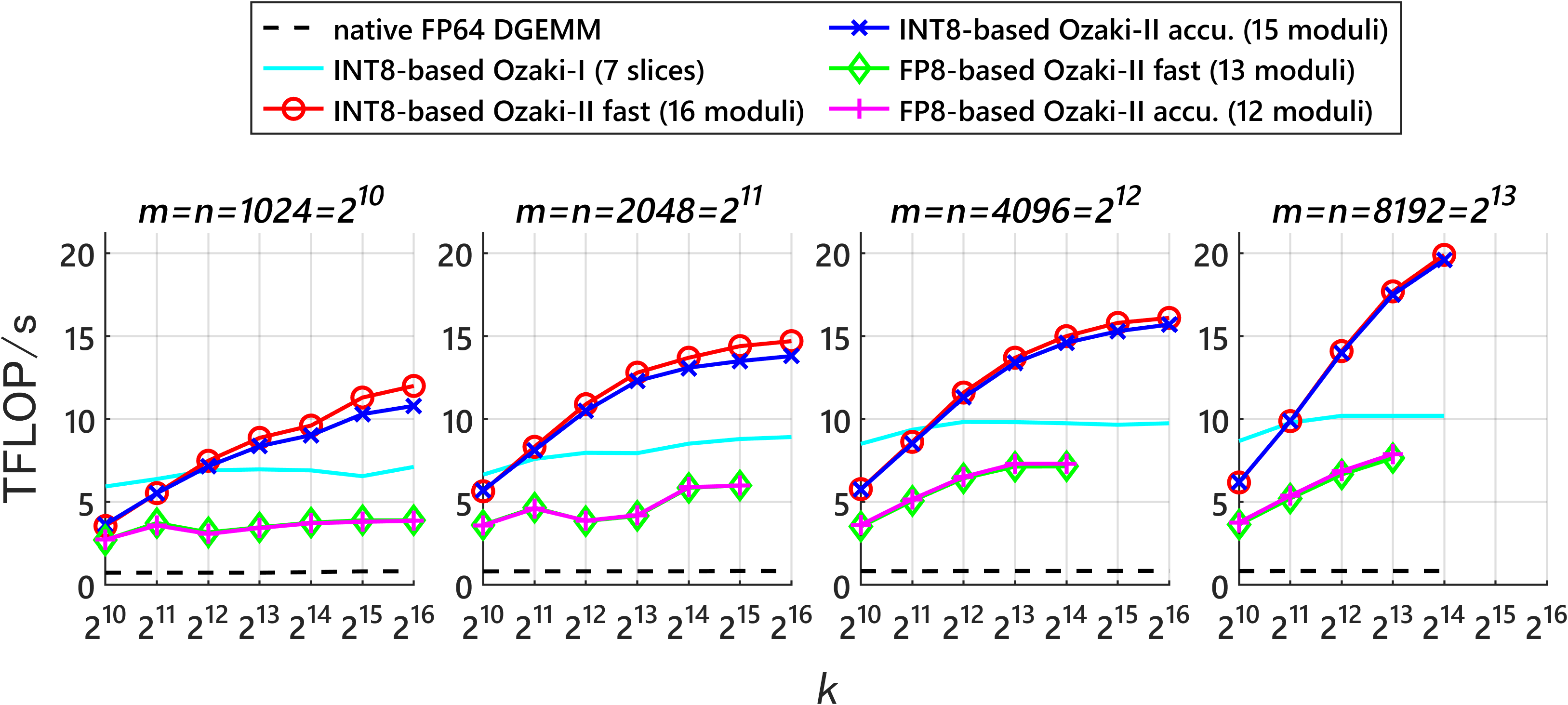}
\caption{Throughput comparison on RTX 5080.\label{fig:flops_RTX5080}}
\end{figure}

\begin{figure}[htbp]
\centering
\includegraphics[width=\hsize]{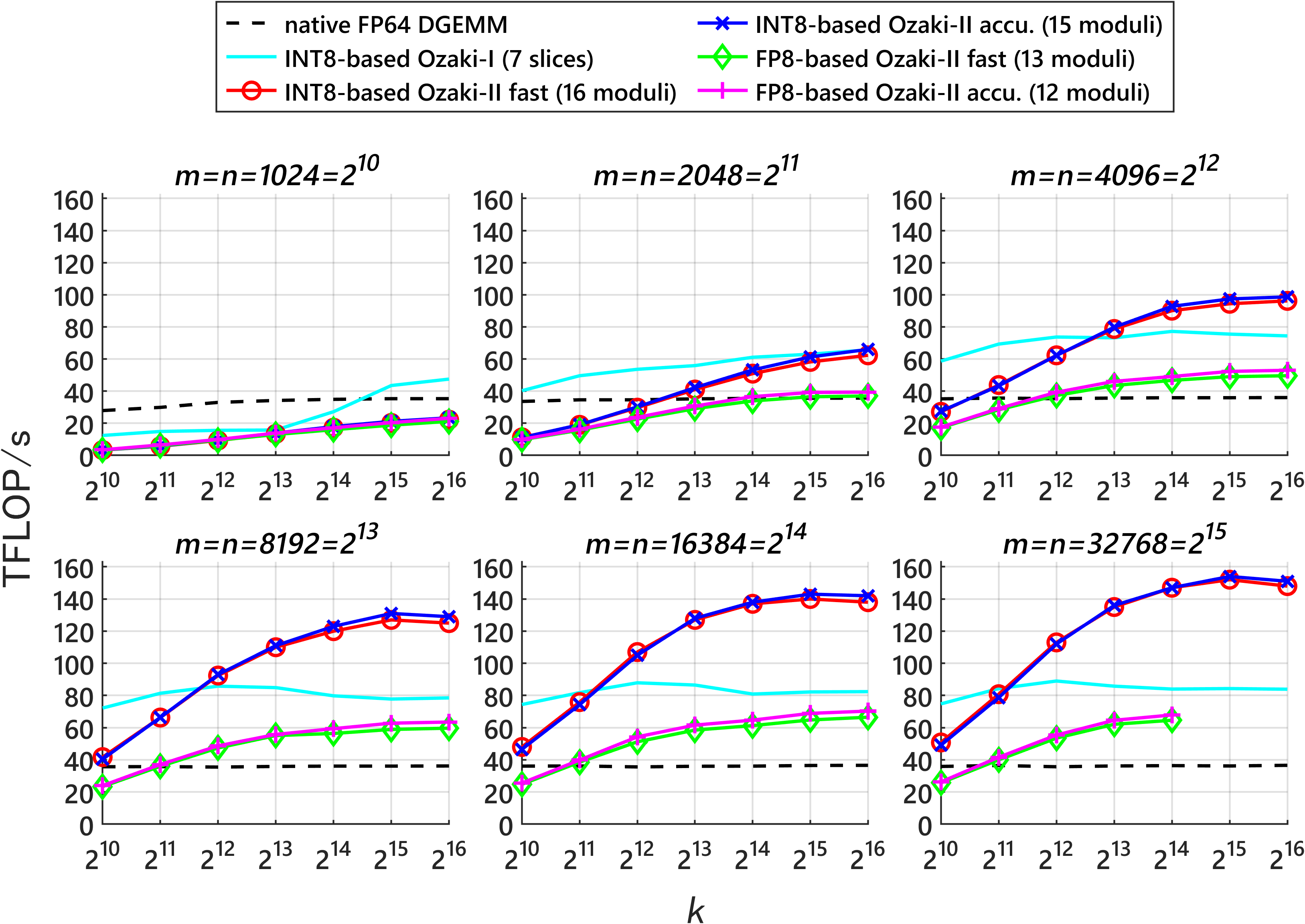}
\caption{Throughput comparison on B200.\label{fig:flops_B200}}
\end{figure}

To examine the dependence on rectangular problem shapes in more detail, Figs.~\ref{fig:flops_RTX5080} and~\ref{fig:flops_B200} present the throughput results on the RTX~5080 and B200, respectively.
On the RTX~5080, the INT8-based Ozaki-II achieves a $4.8$--$15\times$ speedup over native DGEMM for $m=n=1024$, and a $7.4$--$24\times$ speedup for $m=n=8192$. 
The FP8-based Ozaki-II achieves a $3.7$--$5.0\times$ speedup for $m=n=1024$ and a $4.3$--$9.4\times$ speedup for $m=n=8192$. 
The INT8-based Ozaki-II is faster than the FP8-based Ozaki-II by a factor of $1.3$--$3.1\times$ on the RTX~5080 across the evaluated parameter ranges.

On the B200, for $m=n=1024$, both Ozaki-II schemes underperform native DGEMM: the INT8-based Ozaki-II attains $0.1$--$0.7\times$ of native DGEMM throughput, and the FP8-based Ozaki-II attains $0.1$--$0.7\times$ of native DGEMM throughput. 
For $m=n=16384$, the INT8-based Ozaki-II achieves a $1.3$--$3.9\times$ speedup over native DGEMM, and the FP8-based Ozaki-II achieves a $0.7$--$1.9\times$ speedup. 
The INT8-based Ozaki-II is faster than the FP8-based Ozaki-II by a factor of $0.9$--$2.3\times$ across the evaluated parameter ranges.

Across all tested $(m,n,k)$ configurations on the RTX~5080, both the INT8-based and FP8-based Ozaki-II achieve higher throughput than that of native FP64 DGEMM.
On the B200, for $m=n=2048$, the INT8-based Ozaki-II outperforms native FP64 DGEMM for $k > 4096$, whereas the FP8-based Ozaki-II does so for $k\ge 16384$. 
For $m=n=4096$, the INT8-based Ozaki-II outperforms native FP64 DGEMM for $k>1024$, and the FP8-based Ozaki-II does so for $k > 2048$.
These results suggest that $m$-/$n$-blocking can be an effective strategy for reducing the working memory footprint for large-scale problems: by tiling a large DGEMM into subproblems with $m_{\mathrm{blk}}=n_{\mathrm{blk}}\in\{2048,4096\}$ while keeping $k$ unblocked, one can reduce the working memory footprint to that of the blocked subproblems while operating in regimes where the emulation is inferred to outperform native FP64 DGEMM.

At $m=n=k=16384$ on the B200, the INT8-based Ozaki-II attains 137~TFLOP/s in fast mode with 16 moduli and 138~TFLOP/s in accurate mode with 15 moduli. 
In contrast, the FP8-based Ozaki-II attains 61~TFLOP/s in fast mode with 13 moduli and 65~TFLOP/s in accurate mode with 12 moduli.
Since the FP8 matrix multiplication capability and memory bandwidth are essentially the same on a B200 and a B300 (Table~\ref{tab:gpu_peak_performance}), the FP8-based emulation is expected to achieve similar throughput on a B300.
To assess the validity of the performance models, we measured the sustained throughput of the underlying INT8 and FP8 matrix multiplications. 
It was found to be approximately 3~PFLOP/s on the B200. 
In addition, the effective memory bandwidth throughout the workflow was assumed to be 4~TB/s, which is approximately half of the theoretical peak bandwidth. 
The correction parameter $c$ was set, as in Section~\ref{subsec:Performance Modeling and Throughput Comparison}, to the number of low-precision matrix multiplications. 
Substituting these values into the analytic models yields predicted throughput values of 140~TFLOP/s for the INT8-based Ozaki-II in both fast and accurate modes, 69~TFLOP/s for the FP8-based Ozaki-II in fast mode, and 73~TFLOP/s in accurate mode. 
These predictions are in close agreement with the measured values. 
In particular, the model predicts the throughput of the INT8-based Ozaki-II with high accuracy, while also providing reasonable estimates for the FP8-based Ozaki-II. 
These results support the validity of the proposed performance models.

\subsection{Time Breakdown}
\begin{figure}[htbp]
\centering
\includegraphics[width=\hsize]{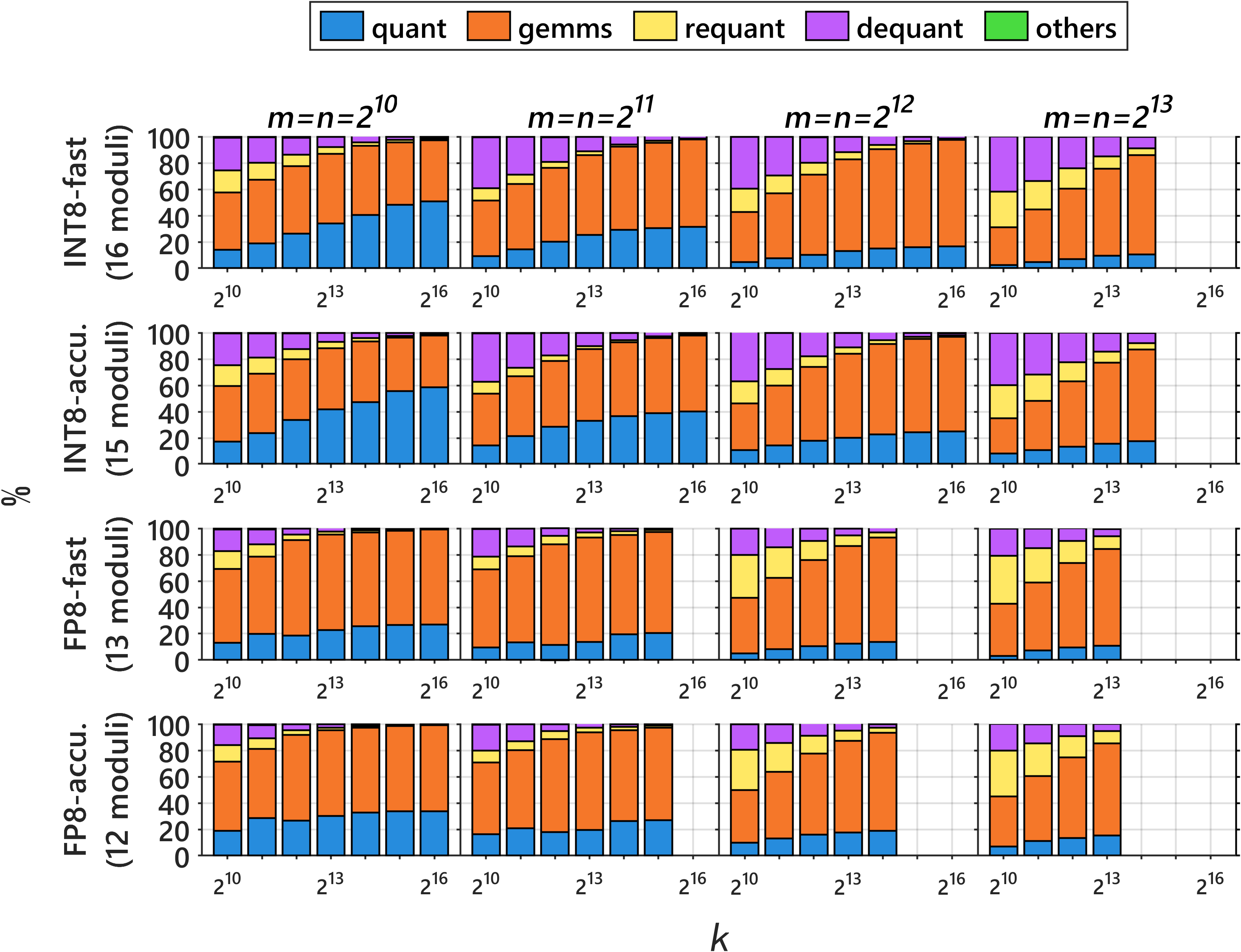}
\caption{Time breakdown for RTX 5080.\label{fig:breakdown_RTX5080}}
\end{figure}

\begin{figure}[htbp]
\centering
\includegraphics[width=\hsize]{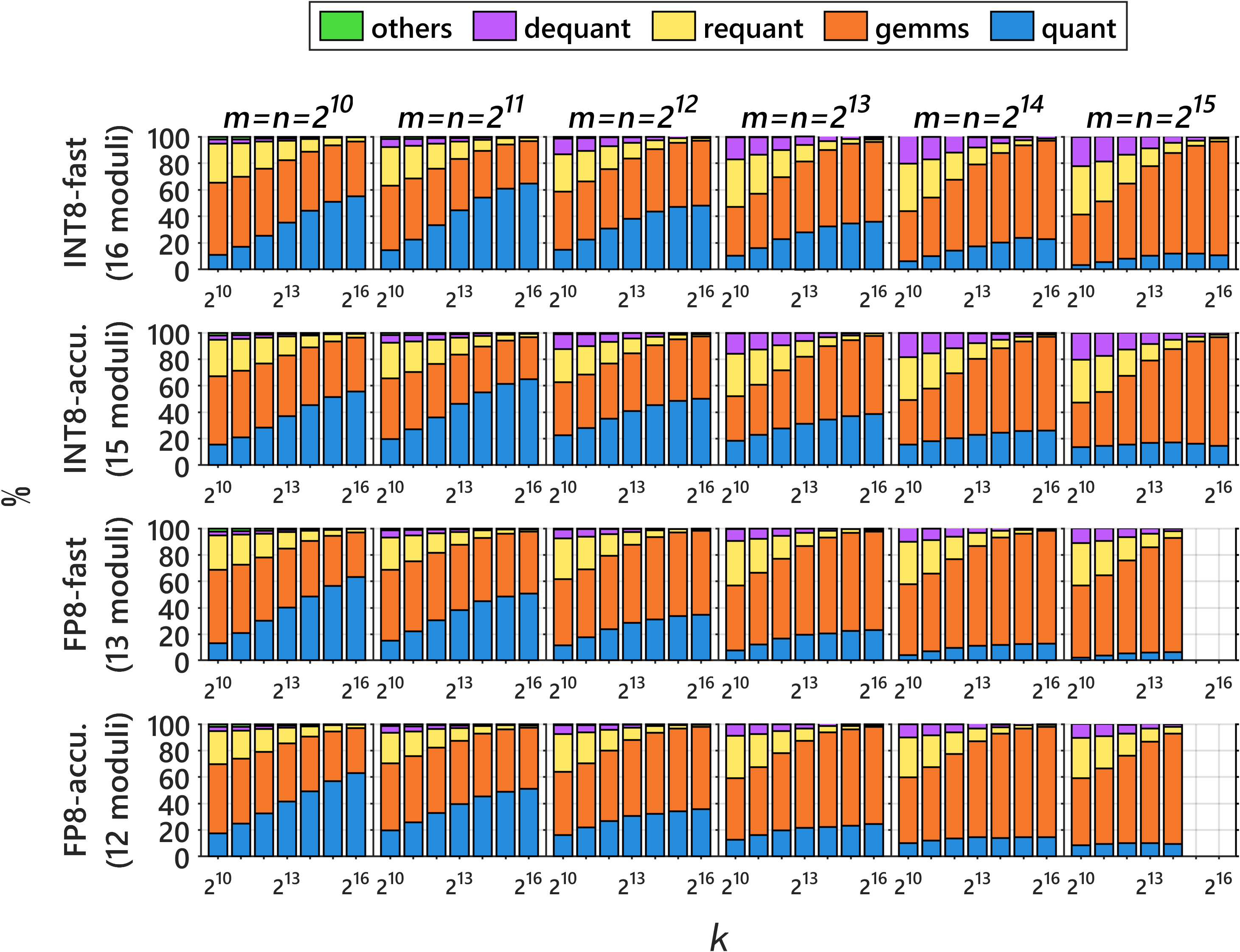}
\caption{Time breakdown for B200.\label{fig:breakdown_B200}}
\end{figure}

We further analyze the runtime composition of the emulation by reporting a GPU time breakdown of the major phases.
We partition the end-to-end time into: 
\begin{description}
    \item[quant] conversion of the input FP64 matrices into INT8/FP8,
    \item[gemms] low-precision matrix multiplications,
    \item[requant] modular reduction of the matrix products,
    \item[dequant] CRT reconstruction and inverse scaling, and
    \item[others] any remaining overhead.
\end{description}
Figs.~\ref{fig:breakdown_RTX5080} and~\ref{fig:breakdown_B200} show the time breakdown for the RTX~5080 and the B200, respectively.
For large $m$ and $n$, the fraction of time spent in \textbf{gemms} increases with $k$, and eventually dominates the end-to-end runtime. 
This is consistent with the transition from a memory-bound regime at small $k$ to a compute-bound regime at larger $k$, where arithmetic throughput of the low-precision GEMM becomes the primary limiter.
For smaller $m$ and $n$, non-\textbf{gemms} phases become comparatively more significant. 
For a small $k$, \textbf{requant} and \textbf{dequant} account for a larger fraction of the runtime, whereas for larger $k$, the \textbf{quant} fraction increases. 
This behavior follows from data-volume scaling: \textbf{requant} and \textbf{dequant} scale primarily with the output size ($mn$), while \textbf{quant} scales with the input sizes ($mk$ and $kn$).
The FP8-based emulation spends a larger fraction of time in \textbf{gemms} than does the INT8-based emulation because each modulus requires three FP8 matrix multiplications, whereas the INT8-based Ozaki-II formulation uses one INT8 matrix multiplication per modulus.

The platform-specific breakdowns further highlight hardware effects. 
On the RTX~5080, the \textbf{dequant} fraction is relatively large, reflecting the limited FP64 arithmetic throughput, since \textbf{dequant} includes FP64-heavy CRT reconstruction and inverse scaling. 
A similar trend was observed on other low-FP64-throughput GPUs, including the RTX~4090~Laptop, RX~9070~XT, and GB10.
On the B200, for small problem sizes, the \textbf{gemms} fraction is smaller than that for the RTX~5080, which is consistent with reduced utilization efficiency of the available MMA units for small matrices.

\section{Conclusion}
\label{sec:Conclusion}
In this paper, we presented an FP8-based DGEMM emulation method based on the Ozaki-II scheme.
By combining a Karatsuba-based extension with a modular reduction technique for selected square moduli, we enabled Ozaki-II-style FP64-targeted emulation on FP8 MMA units and reduced the number of required FP8 matrix multiplications via a hybrid construction.

Through analytic performance modeling, working memory footprint analysis, and empirical benchmarking, we compared the proposed method against INT8-based emulation methods.
Our results indicate that, on platforms where INT8 MMA capacity is sufficient (e.g., INT8 throughput is at least about half of the FP8 throughput), the INT8-based approaches are preferable in terms of both throughput and working memory footprint.
This is a structural consequence of the Ozaki schemes formulation: because the methods transform high-precision matrix multiplication into low-precision fixed-point matrix multiplications, INT8 aligns naturally with the representation, whereas FP8 incurs inefficiency from exponent fields that are not effectively utilized.

Therefore, from the perspective of Ozaki-II-based high-precision matrix-multiplication emulation for HPC, maintaining INT8 capability remains highly beneficial.
Nevertheless, the proposed FP8-based emulation broadens applicability to architectures dominated by low-precision floating-point computation or to hardware with limited INT8 capability, providing a practical option in such environments.
This includes platforms in which INT8 resources are substantially reduced and low-precision floating-point compute dominates (e.g., NVIDIA B300 and NVIDIA Rubin).

\ifanonymous
\else
\section*{Acknowledgment}
The experiments on the B200 were conducted using SAKURA internet Inc.'s managed HPC cluster service, SAKURAONE. We would like to thank Takeshi Yamashita and Fumikazu Konishi of SAKURA internet Inc. for their cooperation in conducting the experiments on the B200.
This study was supported by the Japan Society for the Promotion of Science through Grants-in-Aid for Scientific Research (B) 25K03126, Scientific Research (A) 26H02503, and Early-Career Scientists 26K21234.
\fi

\bibliographystyle{IEEEtran}
\bibliography{refs}

@InProceedings{mukunoki2020,
author="Mukunoki, Daichi
and Ozaki, Katsuhisa
and Ogita, Takeshi
and Imamura, Toshiyuki",
editor="Sadayappan, Ponnuswamy
and Chamberlain, Bradford L.
and Juckeland, Guido
and Ltaief, Hatem",
title="{DGEMM} Using {Tensor Cores}, and Its Accurate and Reproducible Versions",
booktitle="High Performance Computing",
year="2020",
publisher="Springer International Publishing",
address="Cham",
pages="230--248",
url={https://doi.org/10.1007/978-3-030-50743-5\_12}
}

@article{ozaki2012error,
	title={Error-free transformations of matrix multiplication by using fast routines of matrix multiplication and its applications},
	author={Ozaki, Katsuhisa and Ogita, Takeshi and Oishi, Shin'ichi and Rump, Siegfried M.},
	journal={Numerical Algorithms},
	volume={59},
	number={1},
	pages={95--118},
	year={2012},
	publisher={Springer},
    doi = {https://doi.org/10.1007/s11075-011-9478-1},
    url = {https://doi.org/10.1007/s11075-011-9478-1}
}

@article{ozaki2013generalization,
  title={Generalization of error-free transformation for matrix multiplication and its application},
  author={Ozaki, Katsuhisa and Ogita, Takeshi and Oishi, Shin'ichi and Rump, Siegfried M},
  journal={Nonlinear Theory and Its Applications, IEICE},
  volume={4},
  number={1},
  pages={2--11},
  year={2013},
    doi = {https://doi.org/10.1587/nolta.4.2},
    url = {https://doi.org/10.1587/nolta.4.2},
  publisher={The Institute of Electronics, Information and Communication Engineers}
}

@article{ootomo2024dgemm,
author = {Hiroyuki Ootomo and Katsuhisa Ozaki and Rio Yokota},
title ={{DGEMM} on integer matrix multiplication unit},
journal = {The International Journal of High Performance Computing Applications},
volume = {38},
number = {4},
pages = {297-313},
year = {2024},
doi = {10.1177/10943420241239588},
    url = {https://doi.org/10.1177/10943420241239588},
}

@article{uchino2025Performance,
author = {Yuki Uchino and Katsuhisa Ozaki and Toshiyuki Imamura},
title ={Performance enhancement of the {Ozaki Scheme} on integer matrix multiplication unit},
journal = {The International Journal of High Performance Computing Applications},
volume = {39},
number = {3},
pages = {462--476},
year = {2025},
doi = {10.1177/10943420241313064},
url = {https://doi.org/10.1177/10943420241313064},
}

@misc{ozaki-scheme2,
      title={{Ozaki Scheme II}: A {GEMM}-oriented emulation of floating-point matrix multiplication using an integer modular technique}, 
      author={Katsuhisa Ozaki and Yuki Uchino and Toshiyuki Imamura},
      year={2025},
      eprint={2504.08009},
      archivePrefix={arXiv},
      primaryClass={cs.MS},
      url={https://arxiv.org/abs/2504.08009}, 
}

@inproceedings{uchino_ozaki2,
    author = {Uchino, Yuki and Ozaki, Katsuhisa and Imamura, Toshiyuki},
    title = {High-Performance and Power-Efficient Emulation of Matrix Multiplication using {INT8} Matrix Engines},
    year = {2025},
    isbn = {9798400718717},
    publisher = {Association for Computing Machinery},
    address = {St. louis, MO, USA},
    url = {https://doi.org/10.1145/3731599.3767539},
    doi = {10.1145/3731599.3767539},
    booktitle = {Proceedings of the SC '25 Workshops of the International Conference for High Performance Computing, Networking, Storage and Analysis},
    pages = {1824-1831},
    numpages = {8},
    series = {SC Workshops '25}
}

@misc{uchino_ozaki2_complex,
      title={Emulation of Complex Matrix Multiplication based on the {Chinese Remainder Theorem}}, 
      author={Yuki Uchino and Qianxiang Ma and Toshiyuki Imamura and Katsuhisa Ozaki and Patrick Lars Gutsche},
      year={2025},
      eprint={2512.08321},
      archivePrefix={arXiv},
      primaryClass={cs.DC},
      url={https://arxiv.org/abs/2512.08321}, 
}

@inproceedings{mukunoki2025dgemmfp64arithmetic,
author = {Mukunoki, Daichi},
title = {{DGEMM using FP64 Arithmetic Emulation and FP8 Tensor Cores with Ozaki Scheme}},
year = {2026},
isbn = {9798400723285},
publisher = {Association for Computing Machinery},
address = {Osaka, Japan},
url = {https://doi.org/10.1145/3784828.3785017},
doi = {10.1145/3784828.3785017},
booktitle = {Proceedings of the Supercomputing Asia and International Conference on High Performance Computing in Asia Pacific Region Workshops},
pages = {303–311},
numpages = {9},
location = {
},
series = {SCA/HPCAsiaWS '26}
}

@online{Hopper,
author = {{NVIDIA Corporation}},
title = {{NVIDIA H200 Tensor Core GPU}},
year = {2024},
url={https://resources.nvidia.com/en-us-hopper-architecture/hpc-datasheet-sc23},
note = {retrieved 25 January, 2026}
}

@online{Blackwell,
author = {{NVIDIA Corporation}},
title = {{NVIDIA Blackwell Architecture Technical Brief v2.1}},
year = {2025},
url={https://resources.nvidia.com/en-us-blackwell-architecture},
note = {retrieved 25 January, 2026}
}

@online{Rubin,
author = {{NVIDIA Corporation}},
title = {{NVIDIA Vera Rubin NVL72}},
year = {2026},
url={https://www.nvidia.com/en-us/data-center/vera-rubin-nvl72},
note = {retrieved 25 January, 2026}
}

@online{Intel200H200U,
author = {{Intel Corporation}},
title = {{Intel Core Ultra 200H and 200U
Series Processors, Datasheet, Volume 1 of 2}},
year = {2025},
url={https://edc.intel.com/content/www/jp/ja/design/products-and-solutions/processors-and-chipsets/core-ultra-200h-and-200u-series-processors-datasheet-volume-1-of-2/intel-neural-processing-unit-intel-npu/},
note = {retrieved 25 February, 2026}
}

@online{MI300A,
author = {{Advanced Micro Devices, Inc.}},
title = {AMD INSTINCT MI300X APU},
year = {2025},
url={https://www.amd.com/content/dam/amd/en/documents/instinct-tech-docs/data-sheets/amd-instinct-mi300a-data-sheet.pdf},
note = {retrieved 5 December, 2025}
}

@online{MI300X,
author = {{Advanced Micro Devices, Inc.}},
title = {AMD INSTINCT MI300X ACCELERATOR},
year = {2025},
url={https://www.amd.com/content/dam/amd/en/documents/instinct-tech-docs/data-sheets/amd-instinct-mi300x-data-sheet.pdf},
note = {retrieved 5 December, 2025}
}

@online{MI325X,
author = {{Advanced Micro Devices, Inc.}},
title = {AMD INSTINCT MI325X ACCELERATOR},
year = {2025},
url={https://www.amd.com/content/dam/amd/en/documents/instinct-tech-docs/product-briefs/instinct-mi325x-datasheet.pdf},
note = {retrieved 5 December, 2025}
}

@online{MI350X,
author = {{Advanced Micro Devices, Inc.}},
title = {AMD INSTINCT MI350X GPU},
year = {2025},
url={https://www.amd.com/content/dam/amd/en/documents/instinct-tech-docs/product-briefs/amd-instinct-mi350x-gpu-brochure.pdf},
note = {retrieved 5 December, 2025}
}

@online{MI355X,
author = {{Advanced Micro Devices, Inc.}},
title = {AMD INSTINCT MI355X GPU},
year = {2025},
url={https://www.amd.com/content/dam/amd/en/documents/instinct-tech-docs/product-briefs/amd-instinct-mi355x-gpu-brochure.pdf},
note = {retrieved 5 December, 2025}
}

@online{A100,
author = {{NVIDIA Corporation}},
title = {NVIDIA A100 Tensor Core GPU Architecture v1.0},
year = {2020},
url={https://images.nvidia.com/aem-dam/en-zz/Solutions/data-center/nvidia-ampere-architecture-whitepaper.pdf},
note = {retrieved 5 December, 2025}
}

@online{H100,
author = {{NVIDIA Corporation}},
title = {NVIDIA H100 Tensor Core GPU Architecture v1.04},
year = {2023},
url={https://resources.nvidia.com/en-us-hopper-architecture/nvidia-h100-tensor-c},
note = {retrieved 5 December, 2025}
}

@online{TPUv6e,
author = {{Google Cloud}},
title = {TPU v6e},
year = {2026},
url={https://docs.cloud.google.com/tpu/docs/v6e?hl=en},
note = {retrieved 25 February, 2026}
}

@online{TPU7x,
author = {{Google Cloud}},
title = {TPU7x (Ironwood)},
year = {2026},
url={https://docs.cloud.google.com/tpu/docs/tpu7x?hl=en},
note = {retrieved 25 February, 2026}
}

@misc{GTC2025NVIDIA,
author = {Samuel Rodriguez Bernabeu},
title = {{Energy-Efficient Supercomputing Through Tensor Core-Accelerated Mixed-Precision Computing and Floating-Point Emulation}},
howpublished = {Oral presentation at NVIDIA GTC 2025},
  year         = {2025},
  month        = mar,
  day          = {18},
  location     = {San Jose, CA, USA},
url = {https://www.nvidia.com/en-us/on-demand/session/gtc25-s71487/}
}

@inproceedings{NVIDIA2026DGEMM,
author = {Schwarz, Angelika and Anders, Anton and Brower, Cole and Bayraktar, Harun and Gunnels, John and Clark, Kate and Xu, RuQing G. and Rodriguez, Samuel and Cayrols, Sebastien and Tabaszewski, Pawel and Podlozhnyuk, Victor},
title = {Guaranteed DGEMM Accuracy While Using Reduced Precision Tensor Cores Through Extensions of the Ozaki Scheme},
year = {2026},
isbn = {9798400720673},
publisher = {Association for Computing Machinery},
url = {https://doi.org/10.1145/3773656.3773670},
doi = {10.1145/3773656.3773670},
pages = {91–101},
numpages = {11},
location = {Osaka, Japan},
series = {SCA/HPCAsia '26}
}

@online{SAKURAONE,
author = {{SAKURA internet Inc.}},
title = {{SAKURAONE}: a managed high performance computing cluster},
year = {2026},
url={https://www.sakura.ad.jp/sakuraone/},
note = {retrieved 26 March, 2026}
}

@article{rump2012error,
  title={Error estimation of floating-point summation and dot product},
  author={Rump, Siegfried M},
  journal={BIT Numerical Mathematics},
  volume={52},
  number={1},
  pages={201--220},
  year={2012},
  publisher={Springer},
  url = {https://doi.org/10.1007/s10543-011-0342-4}
}


\end{document}